# Induced spin filtering in electron transmission through chiral molecular layers adsorbed on metals with strong spin-orbit coupling

Joel Gersten[(1)], Kristen Kaasbjerg[(2)] and Abraham Nitzan[(2)]

[(1)] Department of Physics, City College of the City University of New York, New York, NY 10031, U.S.A., jgersten@ccny.cuny.edu

[(2)] School of Chemistry, the Sackler Faculty of Science, Tel Aviv University, Tel Aviv, 69978, Israel, nitzan@post.tau.ac.il

## Abstract

Recent observations of considerable spin polarization in photoemission from metal surfaces through monolayers of chiral molecules were followed by several efforts to rationalize the results as the effect of spin-orbit interaction that accompanies electronic motion on helical, or more generally strongly curved, potential surfaces. In this paper we (a) argue, using simple models, that motion in curved force-fields with the typical energies used and the characteristic geometry of DNA cannot account for such observations; (b) introduce the concept of induced spin filtering, whereupon selectivity in the transmission of the electron *orbital* angular momentum can induce spin selectivity in the transmission process provided there is strong spin-orbit coupling in the substrate; and (c) show that the spin polarizability in the tunneling current as well as the photoemission current from gold covered by helical adsorbates can be of the observed order of magnitude. Our results can account for most of the published observations that involved gold and silver substrates, however recent results obtained with an aluminum substrate can be rationalized within the present model only if strong spin-orbit coupling is caused by the built-in electric field at the molecule-metal interface.

# 1. Introduction

Recent observations[1-3] [4] of spin-selective electron transmission through double-strand DNA monolayers adsorbed on gold substrates have attracted considerable interest stemming from the surprising appearance of an apparently large spin-orbit coupling effect in an environment where such large coupling has not been previously observed. Indeed, the following observations need to be rationalized:

(a)[1] High longitudinal (normal to the surface) spin polarization, up to $A \equiv (\uparrow - \downarrow)/(\uparrow + \downarrow) \sim -60\%$ (the "-" sign indicates that the majority spins are antiparallel to the ejecting electron velocity, that is, pointing towards the surface) is observed in photoelectrons ejected from gold covered by self-assembled monolayer of dsDNA at room temperature, largely independent of the polarization of the incident light. The light ($\hbar\omega = 5.84\,\text{eV}$ with pulse duration ~200 ps) was incident normal to the sample.

(b)[1] Using four different lengths of the dsDNA (26, 40, 50 and 78 base pairs) the spin polarization observed on polycrystalline gold surface appears to increase linearly with molecular length. The ~ -60% polarization was obtained with the 78 base-pair monolayer.

(c)[1] ssDNA monolayers show essentially no spin filtering effect (or rather, a small positive polarization which is barely detectable above the experimental noise).

(d)[1] The observed spin polarization is independent of the final kinetic energy of the ejected electron in the range 0…1.2 eV provided by the ejecting light. (Note that the kinetic energy of the emitted electrons reflects the shift in the metal work function caused by the adsorbed monolayer).

(e)[2] Spin selectivity appears to play a role also in the current voltage response of junctions comprising one or a small number of dsDNA oligomers bridging between a nickel substrate and a gold nanoparticle, probed by a conducting AFM (Platinum coated tip) at room temperature. The voltage threshold for conduction and the conduction itself are sensitive to the direction of magnetization induced in the Ni substrate by an underlying magnet, indicating that transmission through the chiral monolayer is sensitive to the spin of the transmitted electron. The effect disappears when a non-chiral layer is used or when the Ni substrate is replaced by gold.

(f)[2] While the I/V behavior appears in these experiments to depend on the length (number of base pairs) of the DNA bridge, the small number of samples used (26, 40 and 50 base pairs) and the statistical noise that characterize single molecule junctions make it impossible to reach a firm conclusion about length dependence of the asymmetry under reversal of magnetization in the Ni substrate.

(g)[4] Spin selectivity appears also in junctions involving dsDNA adsorbed on silver. Here an open circuit configuration was used and the spin-selective electron transfer across the DNA monolayer was inferred from the magnetic field dependence of the voltage induced between the silver substrate and an underlying magnetized nickel, following electron transfer across the DNA from the silver to an optically excited dye attached on the other side of the DNA molecule. The asymmetry is strongly temperature dependent (increasing at lower temperature) and at room temperature appears to be substantially smaller than that the effects described above on gold, although direct comparison cannot be made because of the different experimental configurations used. As in [1], a linear dependence on the DNA chain length is found.

The account of these observations should be supplemented by the well-known fact that photoemission from surface states of solids such as gold characterized by strong spin-orbit coupling using circularly polarized light shows a marked spin polarization that depends on the light polarization.[5] Photoemission from Au(111) with light incident normal to the surface shows[1] electron spin polarization of $A = -22\%$ whose sign reverses with the orientation of the circularly polarized light. No such asymmetry is observed with linearly polarized light. Earlier observations of the overall (not spin resolved) photoemission[6] or photo-induced transmission[7] through chiral molecules induced by circularly polarized light show yields that depend on the combination of molecular helicity and light polarization. This can be rationalized[3] as consistent with the later observations described above: the different spin components transmit through the chiral molecular layer with different efficiencies and, consequently, the overall transmission of the asymmetric spin distribution photoemitted from the gold under circularly polarized illumination depends on the matching between the molecular helicity and the light polarization.

These experimental observations were followed by several attempts to provide theoretical rationalization of these data. It is natural to suspect the implication of spin orbit coupling in the helical molecule as the source for spin selectivity. Indeed, while the atomic spin orbit coupling in carbon is rather small, as evidenced by the very small effect measured in electronic collisions with chiral molecules in the gas phase[8-10] there are experimental and theoretical indications[11-15] that the curvature and torsion imposed on the electron path in helical

4structures such as carbon nanotubes leads to larger spin-orbit coupling than in their linear or planar counterparts[16] due to the overlap of neighboring carbon p orbitals of different symmetries. It is also interesting to note that measured spin-orbit coupling in carbon nanotubes is found[14, 15] to be considerably larger (~3 meV) than is indicated by tight-binding based theoretical calculations that take curvature into account. It is therefore tempting to associate the observations of Refs. [1-4] with spin filtering ideas such as proposed in Refs. [17] and [18].

Recent theoretical efforts[19-25] have pursued this path in somewhat different ways. The authors of Refs. [19] and [20] have considered spin dependent electron scattering by the helical potential in analogy to earlier gas-phase scattering calculations[8-10], while those of Refs. [21, 23, 24] have focused on band motion in a tight-binding helical chain. In both approaches, rather strong, and in our opinion questionable, assumptions are needed to account for the magnitude of the observed effects: Medina et al[20] invoke the density of scattering centers as a source of magnification, but do it by imposing unphysical normalization on the electron wavefunctions while still considering only a few (usually two) scattering events. Gutierrez and coworkers [21] suggest that the origin of the strong observed effect is a strong internal electric field experienced by the electron moving along the helix axis, but do not support this assumption by actual calculations. Furthermore, Guo and coworkers[23] have argued that the model used in Ref. [21] (electron transmission through a single simple helix) should not yield any spin dependent transmission. Instead Guo and coworkers invoke a more complex model, a double helix with interchain interactions in the presence of dephasing, to get asymmetric spin transmission, still without accounting for the magnitude of the effect. Note that the required dephasing appears to stand in contrast to the observation[4] that the magnitude of spin polarization increases at low temperature. Also, unlike the experimental observation, the spin polarization obtained from these calculations appears to be quite sensitive to the electron energy. Note that these tight-binding calculations do not readily account for the observed chain-length dependence (the calculation in Ref. [23] does come close for particular choices of injection energy and dephasing rate), while in the scattering calculation of Ref. [20] this observation is attributed to incoherent additive accumulation of the scattering probability. Finally, Rai and Galperin have shown that pure spin current can be obtained in such tight-binding models from the combined effects of external AC electromagnetic field and DC magnetic field.[25]

In the present paper we examine the possible contribution of *induced* spin filtering to the transmitted spin polarization observed in Refs. [1-4]. As explained below, such contribution to spin filtering by the helical molecular layer reflects the combined effect of orbital angular momentum filtering that characterizes electron transmission through helical molecules and strong spin-orbit coupling at the metal surface. The latter may reflect the intrinsic spin-orbit coupling property of the substrate, and/or Rashba-type coupling associated with built-in electric field at the molecule-metal interface. For example, in the isolated gold atom the energies of the lowest-lying excited states of electronic configurations $(5d)^9(6s)^2$ $^2D_{3/2}$ and $(5d)^9(6s)^2$ $^2D_{5/2}$, with energies of 1.136 eV and 2.658 eV above the ground state, respectively, reflect a spin-orbit splitting of 1.522 eV that results from the intense electric field in the inner core of the atom which is in turn caused by the large atomic number and the short-range screening of the electric field by the core electrons.[26] In gold metal, band-structure calculations of the partial density of states for the d electrons[27-29] show that the spin-orbit splitting in gold and silver are 2.65 eV and 0.79 eV, respectively. These calculations are in agreement with high-resolution x-ray photoemission measurements.[30]

The mechanism considered is similar in spirit to the mechanism for spin polarized photoemission by circularly polarized light[31]. It is simplest to make the point for photoionization of single atoms. Circularly polarized light couples specific eigenstates of the electronic orbital angular momentum, denoted $(l, m_l)$ for a given quantization axis. In the presence of spin-orbit coupling, the atomic angular momentum eigenstates $(j, m_j)$ correspond to the total angular momentum and its azimuthal projection. Still, the information encoded in the selection rules for coupling between the $(l, m_l)$ states affects the transitions between $(j, m_j)$ states (through the corresponding mixing or Clebsch-Gordan coefficients) so as to affect the spin distribution of the ejected or transmitted electrons. The same argument holds for photoemission, in particular when the electrons originate from relatively narrow bands that maintain to some extent the local atomic symmetry. The orbital angular momentum "filtering" in photoemission by circularly polarized light is thus translated at the metal surface to spin filtering.

The proposed mechanism also have some conceptual similarity to a recent suggestion by Vager and Vager,[32] who argue that curvature induced spin orbit coupling leads to correlation between the spin and orbital currents that results in transmitted spin selectivity in any curved path irrespective of the curvature. In our opinion, this correlation implies that the transmission of up-spin with momentum $k$ and that of down-spin with momentum $k - \delta k$ are equally probable, with $\delta k \to 0$ when the curvature vanishes, so it can account for spin selectivity only by fine-tuning a narrow emitted energy window (see also [33]).

The essence of our proposal is that helical molecules can act similarly to circularly polarized light in affecting angular momentum filtering. This is based on the observation that under suitable conditions electron transfer can have the characteristic of current transfer, [34-36] that is, the transferred electron can carry information about its linear and/or angular momentum. Such a picture was used previously[34] to interpret the observations,[6, 7] already mentioned above, that when electron transfer or transmission induced by circularly polarized light take place through chiral molecules, their efficiencies are larger when the light polarization matches the molecular helicity than when it does not. Similarly, in the present case, opposite angular momentum $(\pm m_l)$ states couple differently to the molecular helix and, provided the substrate surface is characterized by strong SO coupling, this orbital angular momentum filtering translates into spin filtering during the injection process. This picture implies that the spin filtering observed in References [1-3] [4] may reflect the spin-orbit coupling at the metal-molecule interface in addition to any spin filtering in the molecular layer itself.

An immediate consequence of this model is the prediction that the effect will be smaller for interfaces with weaker spin-orbit coupling, which seems to be consistent with the weaker effect found on silver,[4] but not with recent results obtained on Aluminum.[37] It should be kept in mind, however, that Rashba spin-orbit coupling can result from strong interfacial fields at metal-molecule interfaces that in turn depend on the electronic chemical potential difference between the metal and the adsorbate layer and are made stronger because of the short electron screening length in the metal. In this paper we explore other implications of this picture, using several different models for the electron propagation through the molecular environment. We start in Section 2 by considering the effect of SO coupling in the helical molecular structure. We analyze two models for electron transport through a helical structure where the SO coupling is derived from the helical potential, and show that such models cannot account for the observed spin polarization. In Section 3 we introduce and discuss the concept of induced filtering. Sections 4 and 5 consider angular momentum selectivity and the consequent spin filtering for different transmission models: One (Section 4) considers electron transmission through a helical tight-binding chain and the other (Section 5) describes on electron scattering by the molecular helical potential. The first model seems to represent the situation incurred for electron tunneling transmission with energy well below the vacuum level, while the other is more suitable for the description of photoemission, where the electron energy is larger than the vacuum level. We calculate the spin filtering associated with each of these models and compare its properties as compared with the experimental observations. Section 5 concludes.

## 2. Spin orbit coupling induced by motion through the helix

In this Section we analyze the implication of electron motion through the helical structure on its spin evolution caused by the ensuing spin orbit coupling. We find that the predicted effect is small.

While the actual motion of the electron should be obtained by solving the Schrödinger equation under the effect of the electron-molecule coupling, we expect that a reasonable order-of-magnitude estimate can be obtained by considering two limiting cases. In one, the electron is assumed to travel in a 1-dimensional path along the helix. In the other the unperturbed electron is assumed to be a plane wave travelling in the z (axial) direction and to be scattered by the helical potential.

*2a. Spin rotation during helical motion.*

Consider an electron moving along a 1-dimensional helical path embedded in 3-dimensional space. The spin degrees of freedom will be treated quantum mechanically and the translational motion will be treated classically. Denote the helix radius by *a*, the pitch by *p* and the speed along the axis of the helix by $v$ (see Fig. 1). For a right-handed helix the location of the electron as a function of time is

$$\begin{aligned} x &= a\cos\left(\frac{2\pi z}{p}\right) \\ y &= a\sin\left(\frac{2\pi z}{p}\right) \\ z &= vt \end{aligned} \quad (1)$$

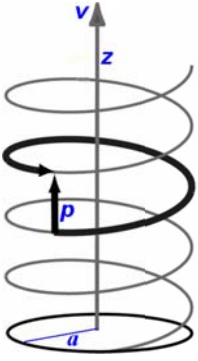

Fig. 1. A generic helical structure referred to in the present section. It is characterized by its radius *a*, pitch *p* and length *L* (the latter not indicated). The transmitting electron is assumed to move with constant speed $v$ along the axial direction.

The velocity and acceleration components are, respectively,

$$\dot{x} = -\frac{2\pi a}{p} v \sin\left(\frac{2\pi z}{p}\right)$$
$$\dot{y} = \frac{2\pi a}{p} v \cos\left(\frac{2\pi z}{p}\right) \quad (2)$$
$$\dot{z} = v$$

and

$$\ddot{x} = -a\left(\frac{2\pi v}{p}\right)^2 \cos\left(\frac{2\pi z}{p}\right)$$
$$\ddot{y} = -a\left(\frac{2\pi v}{p}\right)^2 \sin\left(\frac{2\pi z}{p}\right) \quad (3)$$
$$\ddot{z} = 0$$

By Newton's second law, the force responsible for the acceleration is

$$\vec{F} = m\ddot{\vec{r}} = -\nabla V \quad (4)$$

where *m* is the electron's mass and *V* is the potential confining the electron. Note that this model is simplified in that the presence of discrete molecular groups along the helix is ignored. The spin orbit interaction (including the Thomas 1/2 factor[38]) is given by

$$H_{SO} = -\frac{\hbar}{4mc^2} \vec{\sigma} \cdot \vec{r} \times \nabla V = \frac{\hbar}{4mc^2} \vec{\sigma} \cdot \vec{r} \times \vec{F} \quad (5)$$

Here *c* should be taken to be the speed of light in the adsorbate medium. Thus,

$$H_{SO} = \frac{\hbar}{4c^2}\left(\frac{2\pi v}{p}\right)^2\left[\frac{2\pi v}{p}a^2\sigma_z + av\left(\sigma_x \sin\left(\frac{2\pi vt}{p}\right) - \sigma_y \cos\left(\frac{2\pi vt}{p}\right)\right)\right] \quad (6)$$

These equations are the same in form as those used to describe electron spin resonance or nuclear spin resonance (see e.g. Ref. [39]). The Heisenberg equations of motion for the dynamical Pauli matrices are

$$\frac{d\vec{\sigma}(t)}{dt} = \frac{i}{\hbar}[H_{SO}, \vec{\sigma}(t)] \quad (7)$$

(Other terms in the Hamiltonian commute with $\vec{\sigma}$). Introducing the vector

$$\vec{A}(t) = \frac{\hbar a^2}{4c^2}\left(\frac{2\pi v}{p}\right)^3\hat{k} + \frac{\hbar av}{4c^2}\left(\frac{2\pi v}{p}\right)^2\sin\left(\frac{2\pi vt}{p}\right)\hat{i} - \frac{\hbar av}{4c^2}\left(\frac{2\pi v}{p}\right)^2\cos\left(\frac{2\pi vt}{p}\right)\hat{j} \quad (8)$$

the equations of motion become the familiar Bloch equations for the precession of spin around a time-dependent vector

$$\frac{d\vec{\sigma}(t)}{dt} = \frac{2}{\hbar}\vec{A}(t)\times\vec{\sigma}(t) \quad (9)$$

In terms of the scaled time

$$\theta \equiv \frac{2\pi vt}{p} \quad (10)$$

and the dimensionless parameters

$$b = \frac{\pi a v^2}{pc^2} \quad ; \quad g = \frac{\pi a v}{cp} \quad (11)$$

the equations of motion become

$$\frac{d\sigma_x}{d\theta} = -b\sigma_z\cos\theta - 2g^2\sigma_y$$

$$\frac{d\sigma_y}{d\theta} = -b\sigma_z\sin\theta + 2g^2\sigma_x \quad (12)$$

$$\frac{d\sigma_z}{d\theta} = b(\sigma_y\sin\theta + \sigma_x\cos\theta)$$

In particular, the timescale for changing $\sigma_z$ is seen to be

$$t_{sc} = \left(\frac{2\pi^2 a v^3}{p^2 c^2}\right)^{-1} \quad (13)$$

The corresponding length scale is $z_{sc} = vt_{sc}$ or

$$\frac{z_{sc}}{p} = \frac{p}{2a}\left(\frac{c}{\pi v}\right)^2 \tag{14}$$

Using the parameters $c = 3 \times 10^8$ m/s, $a = 10$ nm, $p = 3.4$ nm, and $v = 5.9 \times 10^5$ m/s (1eV) one gets $z_{sc}/p \sim 4 \times 10^4$. Thus, only little rotation of the spin can be expected when an electron traverses a helix consisting of several turns. This consideration is also supported by a simple perturbation calculation (see Appendix A). The same magnitude of the effect is expected for a cross-coupled double helix structure.

## 2b. *Electron scattering by a helical potential*.

In an alternative picture, consider an electron moving in the outwards z direction while interacting with the helix potential, and the spin polarization that results from the Rashba interaction. Again, the velocity of the electron in the z direction is taken to be constant, so $z = vt$. The motion in the *x*- and *y*- directions will be treated quantum mechanically. The unperturbed Hamiltonian is taken to be

$$H_0 = \frac{p_\perp^2}{2m} + V(x,y,t); \quad \vec{p}_\perp = p_x \hat{i} + p_y \hat{j} \tag{15}$$

where the time dependent helical potential experienced by the electron in its rest frame is modeled as

$$\begin{aligned} V(x,y,t) &= V_0 \delta\left(x\cos\varphi(t) + y\sin\varphi(t) - a\right) \delta\left(-x\sin\varphi(t) + y\cos\varphi(t)\right) \\ &= V_0 \delta(xC + yS - a) \delta(-xS + yC) \end{aligned} \tag{16}$$

where *a* is the radius of the coil, $V_0$ is the strength of the interaction and

$$\varphi(t) = \frac{2\pi vt}{p}, \quad C = \cos\varphi(t), \quad S = \sin\varphi(t) \tag{17}$$

It is convenient to use a rotating coordinates frame by defining

$$\begin{aligned} x' &= x\cos\varphi(t) + y\sin\varphi(t) \equiv xC + yS \\ y' &= -x\sin\varphi(t) + y\cos\varphi(t) \equiv -xS + yC \end{aligned}, \tag{18}$$

in terms of which the model potential is given by

$$V(x',y',t) = V_0 \delta(x'-a) \delta(y'). \tag{19}$$

The magnetic field is given by

$$\vec{B} \approx -\frac{\vec{v} \times \vec{E}}{c^2} = -\frac{1}{ec^2} \vec{v} \times \nabla V \tag{20}$$

where c is the speed of light in the medium. The spin-orbit interaction Hamiltonian is

$$H_{so} = -\vec{\mu} \cdot \vec{B} \tag{21}$$

where $\vec{\mu}$ is the magnetic moment associated with the spin

$$\vec{\mu} = -g\frac{e}{2m}\vec{s} = -g\frac{e}{2m}\frac{\hbar}{2}\vec{\sigma} \tag{22}$$

Eqs. (20)-(22) lead to

$$H_{so} = -\frac{g\hbar}{4mc^2}\vec{\sigma} \cdot \vec{v} \times \nabla V \tag{23}$$

Let $\vec{v} = v\hat{k}$ (where $\hat{k}$ is a unit vector in the direction of the z axis) and take $g = 2$. Introducing also the Thomas factor (1/2), the spin-orbit interaction becomes

$$H_{so} = -\frac{\hbar v}{4mc^2}\vec{\sigma} \cdot \hat{k} \times \nabla V \tag{24}$$

In Appendix B we use time-dependent perturbation theory (first order) to calculate the amplitude for making a transition from an initial state $(\vec{k}_\perp, s)$ to a final state $(\vec{k}'_\perp, s')$, where $\vec{k}_\perp$ corresponds to motion in the *xy*-plane and the spatial parts of the initial and final wavefunctions are $\psi_i = e^{i\vec{k}_\perp \cdot \vec{R}}/\sqrt{A}$ and $\psi_f = e^{i\vec{k}'_\perp \cdot \vec{R}}/\sqrt{A}$, $A$ being the normalization surface in the xy plane (the resulting probability is multiplied below by the number of adsorbed helical molecules in the normalization area, so that the transition cross-section will be proportional to the density $\rho_h$ of such molecules). The result for the transition amplitude is

$$c_{\vec{k}'_\perp, s'} = \frac{ivV_0}{4mc^2}\frac{p}{2\pi v}\frac{1}{A}\int_0^{2\pi N} e^{iqa\cos\varphi} \chi_{s'}^\dagger \begin{pmatrix} 0 & -q_- \\ q_+ & 0 \end{pmatrix} \chi_s d\varphi \tag{25}$$

where $\chi_s$ and $\chi_{s'}$ are the initial and final spin vectors, $\vec{q} = \vec{k}_\perp - \vec{k}'_\perp$ and $q_\pm = q_x \pm iq_y$. In the case that *N* is an integer one obtains

$$c_{\vec{k}'_\perp, s'} = \frac{iNpV_0}{4mc^2}\frac{1}{A}J_0(qa)\chi_{s'}^\dagger \begin{pmatrix} 0 & -q_- \\ q_+ & 0 \end{pmatrix} \chi_s \tag{26}$$

Spin flip transitions occur when $(s,s') = (1,-1)$ or $(-1,1)$. The transition probability in either case is obtained from the square of the amplitude calculated from (26) multiplied by the density of final states, $A/(2\pi)^2$

$$P(q) = \frac{1}{A}\left(\frac{NV_0 pq J_0(qa)}{16\pi mc^2}\right)^2 \rightarrow \rho_h\left(\frac{NV_0 pq J_0(qa)}{16\pi mc^2}\right)^2 \quad (27)$$

The expression on the right is obtained by further multiplying by the number of helical molecules, $N_h = \rho_h A$, adsorbed in the normalization area, where $\rho_h$ is the surface density of such molecules. The transition probability (27) depends on the square of the molecular length. More importantly, there is no difference between positive and negative helicity states and hence no spin selectivity occurs for a given $q$.

The total transition probability is obtained by integrating over all wave-vector transfers

$$P_{total} = \int d^2q P(q) = \frac{1}{2\pi}\frac{(V_0 Np)^2}{(4mc^2)^2}\frac{1}{A}\int_0^{2k} dq\, q^3 J_0^2(qa) =$$
$$\chi \frac{4a^2 k^2}{3}\left[2(ka)^2 J_0^2(2ka) + 2ka J_0(2ka) J_1(2ka) + (2k^2 a^2 - 1) J_1^2(2ka)\right] \quad (28)$$

where the parameter $\chi$ is

$$\chi = \frac{\rho_h}{2\pi}\frac{(V_0 Np)^2}{(4mc^2 a^2)^2} \quad (29)$$

For $V_0 = 10$ eV nm$^2$, $Np = 50$ nm, $mc^2 = 511$ keV (for this estimate we use the speed of light in vacuum), $a = 1$ nm, and $\rho_h = 0.3/\text{nm}^2$ this yields $\chi \approx 10^{-9}$. For an electron energy of 1 eV the value of $k$ is 6x10$^9$ so $ka \approx 6$ and the order of magnitude of the result is not changed by much. Furthermore, the transition probability is symmetric for positive and negative helicities and equivalently, as noted above, for $+ \rightarrow -$ and $- \rightarrow +$ spin transitions. It therefore does not lead to a net spin polarization. In Appendix B we further show that a model with two helical molecules yields essentially similar results. In conclusion, the axial motion of the electron through the helix is not a good model for explaining the spin polarization of the electrons that pass through the DNA molecule.

It should be noted that because motion in the z direction has be taken classical, the calculation that lead to Eq. (28) does not take into account possible constructive interference in the diffraction of the electronic wavefunction from the periodic helix structure (see Section 5,

Eq. (88) and the discussion following it for a treatment that takes this constructive interference into account). Removing this shortcoming still leads to a negligibly small contribution.[40]

## 3. Induced filtering

We define induced filtering (or induced selectivity) as a process where geometry or symmetry-imposed selectivity in one variable $A$ causes selectivity in another variable $B$ that is coupled to it. Specifically, let $\hat{H} = \hat{A} + \hat{B}$ where $[\hat{A}, \hat{B}] = 0$. Then eigenstates of the Hamiltonian can be written as products $\psi_{a,b} = |a\rangle|b\rangle$ of eigenstates of the $\hat{A}$ and $\hat{B}$ operators. When in such a state, the probability to observe the system in eigenstate state $b'$ of $\hat{B}$, is obviously $\delta_{b,b'}$. Consider the transformation $\psi_{a,b} \to \psi_{a',b}$ induced by some external or internal perturbation represented by a hermitian operator $\hat{V}$ that couples only states in the $\{|a\rangle\}$ sub-space, so that $\psi_{a'b} = \hat{V}\psi_{a,b}$ (e.g., in optical transitions $\hat{V}$ is often the dipole operator) with $\|\psi_{a',b}\| = 1$. As indicated, the state in the B subspace is not affected by this coupling, so the probability of observing a particular value of the B variable is the same before and after the transition. Indeed,

$$P_{b'} = \langle b'|\mathrm{Tr}_A(|V\psi_{a,b}\rangle\langle V\psi_{a,b}|)|b'\rangle = \delta_{b,b'}\sum_{a'}\langle a|V|a'\rangle\langle a'|V|a\rangle = \delta_{b,b'} \qquad (30)$$

In the presence of coupling between $\hat{A}$ and $\hat{B}$ the eigenstates of the system Hamiltonian are no longer simple products of $|a\rangle$ and $|b\rangle$, but can be expanded in the form

$$\psi = \sum_a \sum_b c_{ab}|a\rangle|b\rangle \qquad (31)$$

For a system in this state the reduced density matrix in the B subspace is

$$\rho_{b,b'}^{(B)} = \mathrm{Tr}_A(|\psi\rangle\langle\psi|) = \sum_a c_{ab}c_{ab'}^* \qquad (32)$$

and in particular, the probability to observe the system in state $b$ is

$$P_b = \sum_a |c_{ab}|^2 \qquad (33)$$

The transformed state is now $\hat{V}\psi = \sum_a \sum_b c_{ab} |b\rangle \hat{V}|a\rangle$ and the reduced density matrix in the B subspace is

$$\rho_{b,b'}^{(B)} = \mathrm{Tr}_A\left(\hat{V}|\psi\rangle\langle\psi|\hat{V}\right) = \sum_{a''}\sum_{a,a'} c_{ab} c_{a'b'}^* \langle a''|V|a\rangle \langle a'|V|a''\rangle$$
$$= \sum_{a,a'} c_{ab} c_{a'b'}^* \langle a'|VV|a\rangle \quad (34)$$

The probability to observe $b$ is then

$$P_b = \sum_{a,a'} c_{ab} c_{a'b}^* \langle a'|VV|a\rangle \quad (35)$$

Comparing to (33) we see that the final distribution in the B-subspace is now affected by the transition - any selectivity expressed by the $\langle a'|VV|a\rangle$ matrix elements is partly imparted into this distribution. As a variation of this theme we note that instead of $\mathrm{Tr}_A$ in (34) we often need to sum only over states on an energy shell, so that the quantity of interest is

$$\rho_{b,b'}^{(B)}(E) = \sum_{a''}\sum_{a,a'} c_{ab} c_{a'b'}^* \langle a''|V|a\rangle \langle a'|V|a''\rangle \delta(E - E_{a''})$$
$$= \frac{1}{2\pi} \sum_{a,a'} c_{ab} c_{a'b'}^* \Gamma_{a,a'}(E) \quad (36)$$

where

$$\Gamma_{a,a'}(E) = 2\pi \sum_{a''} \langle a'|V|a''\rangle \langle a''|V|a\rangle \delta(E - E_{a''}) \quad (37)$$

Eqs. (35) and (36) are manifestations of induced selectivity. In what follow we consider two concrete examples.

*Induced selectivity in transmission.* Consider a junction in which a molecular bridge $M$ connects two free electron reservoirs, $L$ (left) and $R$ (right), as seen in Fig. 2. The transmission function $\mathcal{T}(E)$ is given by the Landauer formula

$$\mathcal{T}(E) = \mathrm{Tr}\left[\Gamma^{(L)}(E) G^\dagger(E) \Gamma^{(R)}(E) G(E)\right] \quad (38)$$

where $G$ is the molecular retarded Green function and $\Gamma^{(K)}$, $K = L, R$ is twice the imaginary part of the self-energy of the bridge associated with its coupling to the reservoir $K$ and the trace is over the bridge subspace. In the basis of eigenstates of the bridge Hamiltonian

$$\left(\Gamma^{(K)}(E)\right)_{n,n'} = 2\pi \sum_k V_{nk} V_{k,n'} \delta(E - \varepsilon_k), \tag{39}$$

where the sum is over the free electron states of energies $\varepsilon_k$ in the reservoir K. The subscript k enumerates the states in the reservoirs, and is usually associated with eigenstates of operators that appear in the Hamiltonian and commute with it. Consider now the situation where the single electron states in R are characterized, in addition to their energy, by quantum numbers $l$, $s$ associated with independent operators $\hat{L}$ and $\hat{S}$ that commute with the Hamiltonian, while in L these operators are coupled by some internal single electron force field, so that only some combined operator $\hat{J}$ commutes with the Hamiltonian. For example, if the left and right electronic reservoirs are metals with single electron states described by Bloch wavefunctions $\psi_{n\mathbf{k}}(\mathbf{r}) = e^{i\mathbf{k}\cdot\mathbf{r}} u_{n\mathbf{k}}(\mathbf{r})$, the quantum numbers $n$ (or a set of such numbers) characterizing different bands will have atomic character if the bands are narrow relative to the spacing between the parent atomic levels. In such a case, $n$ can stand for the quantum numbers $(l,s)$ of the orbital and spin angular momenta in a metal with no spin-orbit coupling, while in the presence of spin orbit coupling only the state $j$ of the total angular momentum is meaningful. (Note that in reality we should also consider the projections of these vector operators on some axis, and the corresponding quantum numbers $m_l, m_s$ and $m_j$. This is done in the application discussed in Section 4). Equation (39) can then be recast in more detailed forms

$$\left(\Gamma^{(L)}(E)\right)_{n,n'} = \sum_j \Gamma^{(Lj)}; \quad \left(\Gamma^{(Lj)}\right)_{n,n'} = 2\pi \sum_k V_{n,jk} V_{jk,n'} \delta(E - \varepsilon_{jk}) \tag{40a}$$

$$\left(\Gamma^{(R)}(E)\right)_{n,n'} = \sum_{l,s} \Gamma^{(Rls)}; \quad \left(\Gamma^{(Rls)}\right)_{n,n'} = 2\pi \sum_k V_{n,lsk} V_{lsk,n'} \delta(E - \varepsilon_{lsk}) \tag{40b}$$

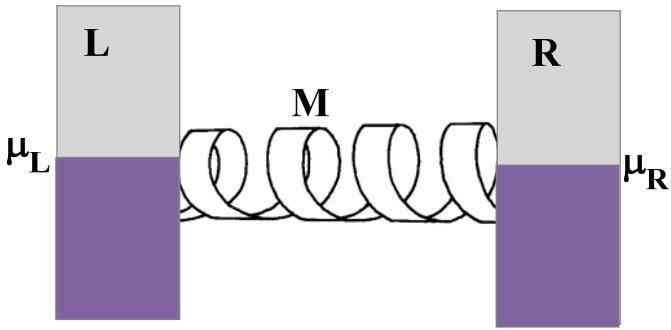

Fig. 2. A schematic view of a helical molecular bridge connecting two metal leads, left and right, characterized by the electronic electrochemical potentials, $\mu_L, \mu_R$, respectively.

Eqs. (40) are identical to what will be obtained in a multi-terminal junction, where each of the $\{Lj\}$ and $\{Rls\}$ groups of states represent different terminals. The transmission fluxes between two such terminals take the forms[41]

$$\mathcal{T}_{Lj \to Rls}(E) = \text{Tr}\left[\Gamma^{(Lj)}(E) G^\dagger(E) \Gamma^{(Rls)}(E) G(E)\right] \quad (41)$$

Whether such fluxes are measurable or not depend on the energetic details of the system. For example, if the bands *j* are energetically distinct, it is possible in principle, by tuning the voltage bias window, to focus on the flux associated with a particular "*j* terminal". If the electrons emerging on the right can be also analyzed and their quantum state (*l,s*) can be determined, we are in a position to determine the flux associated with the transmission function of Eq. (41).

In the application considered in Section 4 we are interested in the transmission into a particular eigenstate *s* of the operator $\hat{S}$ (that is, in an experiment where $\hat{S}$ is monitored in the outgoing electronic flux in *R*). This corresponds to the transmission function

$$\mathcal{T}_{Lj \to Rs}(E) = \sum_l \mathcal{T}_{Lj \to Rls}(E) = \text{Tr}\left[\Gamma^{(Lj)}(E) G^\dagger(E) \Gamma^{(Rs)}(E) G(E)\right]$$
$$\Gamma^{(Rs)} \equiv \sum_l \Gamma^{(Rls)} \quad (42)$$

Suppose now that the bridge Hamiltonian, as well as the Hamiltonian of the *R* reservoir and the couplings between the bridge and the reservoirs, do not depend on the operator $\hat{S}$. In this case (cf. Eq. (40)) $\left(\Gamma^{(Rls)}\right)_{n,n'} = 2\pi \sum_k V_{n,lk} V_{lk,n'} \delta(E - \varepsilon_{lk}) \equiv \Gamma^{(Rl)}$ as well as $\Gamma^{(Rs)}$ defined in Eq. (42) do not depend on *s*. In particular, $\Gamma^{(Rs)}$ will be denoted $\Gamma^{(R)}$ below. On the left, writing (as in (31)),

$$|j\rangle = \sum_l \sum_s c_{j,ls} |l\rangle |s\rangle \quad (43)$$

we find

$$\left(\Gamma^{(Lj)}\right)_{n,n'} = 2\pi \sum_k \sum_l \sum_{l'} \sum_s c_{j,ls} c^*_{j,l's} V_{n,lk} V_{l'k,n'} \delta(E - \varepsilon_{jk}) = \sum_s \Gamma^{(Ljs)} \quad (44a)$$

$$\left(\Gamma^{(Ljs)}\right)_{n,n'} \equiv 2\pi \sum_k \sum_l \sum_{l'} c_{j,ls} c_{j,l's}^* V_{n,lk} V_{l'k,n'} \delta\left(E - \varepsilon_{jk}\right) \tag{44b}$$

If in addition we disregard in the Green functions in Eq. (42) terms that make them non spin-diagonal, then the separability of Eq. (44) into its *s* components make it possible to write the transmission function for the flux from the *L* terminal into a particular state *s* in the outgoing flux on the right in the form

$$\mathcal{T}_{Lj \to Rs}(E) = \text{Tr}\left[\Gamma^{(Ljs)}(E) G^\dagger(E) \Gamma^{(R)}(E) G(E)\right] \tag{45}$$

Two comments regarding this result are in order: First, although we have made the assumption the bridge Hamiltonian does not depend on *S*, the Green functions $G(E)$ are in general non spin-diagonal because of the self-energy terms associated with the coupling between the bridge and the left reservoir in which strong S-L coupling exists. Such terms can couple different *s* states of the bridge through their interaction with the same *j*-state on the left reservoir. Such couplings have been disregarded in obtaining Eq. (45) - a reasonable approximation when the molecule-lead coupling is not too strong. Second, the appearance of the subscript *j* reflects our assumption that transmission out of the *L* reservoir is dominated by a particular band whose atomic origin is indicated by the quantum number *j*.

Equation (45) constitutes our final result for this case. To see its significance consider, for example, the transmission from *L* to *R* as would be realized if the bias is such that (a) a particular band *j* in *L* is the source, and (b) the *L* states are occupied while their *R* counterparts are vacant. The probability to measure a value *s* for the observable $\hat{S}$ in the source terminal is, from Eq. (43)

$$P_s^{(Lj)} = \sum_l \left|c_{j,sl}\right|^2 \tag{46}$$

while the probability for this measurement in the exit terminal is given by

$$P_s^{(R)} = \frac{\mathcal{T}_{Lj \to Rs}(E)}{\sum_s \mathcal{T}_{Lj \to Rs}(E)} \tag{47}$$

Obviously, $P_s^{(R)} \neq P_s^{(Lj)}$, implying that the bridge acts as a $\hat{S}$ filter although its transmission properties do not depend on $\hat{S}$. As already noted, in Section 4 we will replace *j*, *l*, and *s* by $(j, m_j)$, $(l, m_l)$ and $(s, m_s)$ - the quantum numbers that characterize the total, orbital and spin atomic angular momenta and their projections, respectively.

Finally, we note two simplified special cases. First, in the case of a single state bridge, or when the coupling between the bridge and the left terminal is channeled through a single state of the bridge (denoted by $|1\rangle$), Eqs. (47) and (45) lead to

$$P_s^{(R)} = \frac{\Gamma^{(Ljs)}}{\sum_s \Gamma^{(Ljs)}}; \quad \Gamma^{(Ljs)} \equiv 2\pi \sum_k \sum_l \sum_{l'} c_{j,ls} c^*_{j,l's} V_{1,lk} V_{l'k,1} \delta(E - \varepsilon_{jk}) \quad (48)$$

Second, as will be seen below, sometimes the sum over $l,l'$ is dominated by the diagonal $l = l'$ contributions, in which case

$$\Gamma^{(Ljs)} = 2\pi \sum_k \sum_l |c_{j,ls}|^2 |V_{1,lk}|^2 \delta(E - \varepsilon_{jk}) \quad (49)$$

*Induced selectivity in photoemission.* We consider photoemission from a simple atomic lattice model, where the electronic bands are narrow relative to the energy separation between the electronic levels of the constituent atoms. Photoemission then reflects the symmetry property of ionization from a single atom with one difference - the existence of the solid-vacuum interface. Accordingly, we consider a one-electron atom located at the origin and positioned at a distance *a* to the left of this interface, represented by a planar surface, $z = a$. To the right of this surface is vacuum. The interface is simply treated as a step potential given by

$$V(z) = \begin{cases} V_0 & \text{if } z < a \\ 0 & \text{if } z > a \end{cases} \quad (50)$$

where $V_0 < 0$.

Consider next the atomic state in the absence of the interfacial wall. It is taken to be an eigenstate of total angular momentum operator $\hat{j}$ and its azimuthal projection $\hat{j}_z$, with quantum numbers *j* and $m_j$, respectively. In terms of eigenstates of the orbital angular momentum and spin operators, the corresponding wavefunction takes the form

$$u_{j,m_j}(\vec{r}) = \sum_{m_l, m_s} \langle lm_l sm_s | jm_j \rangle v_{j,l}(r) Y_{l,m_l}(\theta, \varphi) \chi^s_{m_s} \quad (51)$$

Here $Y_{l,m_l}(\theta,\varphi)$ are eigenfunctions of the angular momentum operator, $v_{j,l}(r)$ are their radial counterparts and $\chi^s_{m_s}$ are spin wavefunctions - two-component spinors. The symbols $\langle lm_l sm_s|jm_j\rangle$ are Clebsch-Gordan coefficients. We assume that this wavefunction was obtained by absorbing a photon, so its energy $E$ is positive and the state is $2j+1$ degenerate.[42]

This atomic wavefunction is embedded in a continuum of states associated with the semi-infinite spaces to the right and left of the wall at $z = a$. In the vacuum, $z > a$, the Schrödinger equation is

$$\left(\nabla^2 + k^2\right)\phi_R(R,\varphi,z) = 0 \; ; \; k^2 = 2mE/\hbar^2 \tag{52}$$

where $m$ is the (effective) electron mass. The relevant solutions may be expressed in the form of a sum of Bessel transforms

$$\phi_R(R,\varphi,z) = \sum_{m_l,m_s}\int_0^\infty A_{m_l,m_s}(Q) J_{m_l}(QR) e^{iqz+im_l\varphi} \chi^s_{m_s} dQ \tag{53}$$

where $q = \sqrt{k^2 - Q^2}$ for $Q < k$ and $q = i\sqrt{Q^2 - k^2}$ for $Q > k$. In the former case these are outgoing waves into vacuum whereas the latter case describes evanescent waves in the vacuum side of the interface.

In the solid, $z < a$, the free Schrodinger equation is

$$\left(\nabla^2 + \kappa^2\right)\phi_L(R,\varphi,z) = 0 \tag{54}$$

where $\kappa = \sqrt{k^2 - 2m|V_0|/\hbar^2}$. The outgoing solution, i.e., a left-travelling wave, may also be expressed as a sum of Bessel transforms

$$\phi_L(R,\varphi,z) = \sum_{m_l,m_s}\int_0^\infty B_{m_l,m_s}(Q) J_{m_l}(QR) e^{-iq'z+im_l\varphi} \chi^s_{m_s} dQ \tag{55}$$

where $q' = \sqrt{\kappa^2 - Q^2}$.

We will approximate the total wave function in the solid with the atom as a linear combination of an atomic wave function and the free wave function. By using the free wave function rather than the more general solution of the solid-plus-atom potential we are neglecting final-state interactions.[43] In this approximation, the total wave function for $z < a$ is

$$\phi(R,\varphi,z) = \phi_L(R,\varphi,z) + u_{j,m_j}(\vec{r}) \tag{56}$$

whereas in vacuum, $z > a$, the total wave function is simply

$$\phi(R,\varphi,z) = \phi_R(R,\varphi,z) \tag{57}$$

This wavefunction represents a scattering 'in-state'. A similar construction may be used to obtain the scattering 'out-state'.

The expansion coefficients *A* and *B* in (53) and (55) can be found by matching the wave functions at the surface $z = a$, using the continuity of $\phi$ and its normal derivative at this surface. This is done in Appendix C, leading to

$$\begin{bmatrix} A_{m_l,m_s}(Q) \\ B_{m_l,m_s}(Q) \end{bmatrix} = \frac{1}{q+q'}\begin{bmatrix} q'e^{-iqa} & -ie^{-iqa} \\ -qe^{iq'a} & -ie^{iq'a} \end{bmatrix}\begin{bmatrix} \alpha \\ \beta \end{bmatrix} \tag{58}$$

where

$$\alpha = Q\langle lm_l sm_s | jm_j \rangle \int_0^\infty J_{m_l}(QR) v_{j,l}\left(\sqrt{R^2+a^2}\right) Y_{l,m_l}(\Theta,0) R\,dR \tag{59}$$

and

$$\beta = Q\langle lm_l sm_s | jm_j \rangle \int_0^\infty J_{m_l}(QR)\left(\left[\cos\theta\frac{\partial}{\partial r} - \frac{\sin\theta}{r}\frac{\partial}{\partial \theta}\right]v_{j,l}(r) Y_{l,m_l}(\theta,0)\right)_{\substack{r=\sqrt{R^2+a^2}\\ \theta=\Theta}} R\,dR \tag{60}$$

The coefficients $A_{m_l,m_s}(Q)$ and $B_{m_l,m_s}(Q)$ are seen to be simply proportional to the Clebsch-Gordan coefficients.

Eqs. (53), (57), and (58)-(60) give an explicit expression for the outgoing solution outside the solid. The emitted electron flux in the direction normal to the surface is

$$J_z = \frac{\hbar}{2mi}\left(\phi_R^* \frac{\partial}{\partial z}\phi_R - \phi_R \frac{\partial}{\partial z}\phi_R^*\right) \tag{61}$$

The total current is obtained by integrating the current density over an area in vacuum parallel to the surface

$$I_z = \int_0^\infty dR R \int_0^{2\pi} d\varphi J_z = \frac{2\pi\hbar}{m} \sum_{m_l,m_s} \int_0^k dQ \frac{q}{Q} |A_{m_l,m_s}(Q)|^2$$

$$= \frac{2\pi\hbar}{m} \sum_{m_l,m_s} |\langle l m_l s m_s | j m_j \rangle|^2 \int_0^k dQ \frac{qQ}{(q+q')^2} \left[ q'^2 \left( \int_0^\infty J_{m_l}(QR) v_{j,l}\left(\sqrt{R^2+a^2}\right) Y_{l,m_l}(\Theta,0) R dR \right)^2 \right.$$

$$\left. + \left( \int_0^\infty J_{m_l}(QR) \left( \left[\cos\theta \frac{\partial}{\partial r} - \frac{\sin\theta}{r}\frac{\partial}{\partial \theta}\right] v_{j,l}(r) Y_{l,m_l}(\theta,0) \right)_{\substack{r=\sqrt{R^2+a^2}\\\theta=\Theta}} R dR \right)^2 \right] \quad (62)$$

To get (62) we have used the identities

$$\int_0^\infty e^{i(m_l - m_l')\varphi} d\varphi = 2\pi \delta_{m_l,m_l'} \quad (63)$$

and (orthogonality for spinors)

$$\chi_{m_s'}^{s,\dagger} \chi_{m_s}^s = \delta_{m_s,m_s'} \quad (64)$$

Also, the upper limit of the $Q$-integration has been changed from $\infty$ to $k$ since for $Q > k$ the variable $q$ is imaginary and there is no contribution to the current.

As before (Eqs. (45) and (44b)), the appearance here of the Clebsch-Gordan coefficients in the emitted current implies that if an $m_l$-filter was in effect, induced filtering of $m_s$ could result. In particular, $m_l$ selectivity can be imposed by circularly polarized light. Indeed it should be noted that our treatment is an analogue of the Fano theory of spin-polarized photoemission from atoms characterized by strong spin-orbit coupling,[31] generalized to the presence of the solid-vacuum interface.

**4. Induced spin filtering in tunneling through a molecular helix**

Here we implement the results from Section 3, Eqs. (44), (45) and (47), to calculate the induced spin selectivity in a model that incorporates a metal substrate and an adsorbed helical molecule. While we keep the calculation at a generic level, we use the band structure of gold and the structure of the DNA helix to choose specific parameters when needed. It should be emphasized that the actual behavior of electron tunneling between substrate and adsorbate depends on details of the electronic structure as manifested mainly in the alignment between their levels and in their electronic coupling. As we are not using such data but instead make assumptions and take shortcuts in order to simplify the calculation,[44] the results obtained below should be regarded as suggestive of the order of magnitude of the spin polarization effect, rather than conclusive. In order to get such estimate, the following assumptions are made

(a) The tunneling electrons originate primarily from the relatively narrow d-band of gold. More specifically, this band split into a higher energy $^2D_{5/2}$ and a lower energy $^2D_{3/2}$ band which are somewhat overlapping,[27] and we assume that the tunneling current is dominated by the $^2D_{5/2}$ sub-band. This spectroscopic term reflects the atomic parenthood of these states, of orbital angular momentum $l = 2$ and total angular momentum $j = 5/2$.

(b) The DNA molecule is represented by a tight binding helical chain with nearest neighbor intersite coupling $V$ and axis normal to the gold surface, taken below as the z direction.[45]

(c) The DNA-substrate coupling is dominated by the substrate atom at position $\vec{r}_A$ nearest to the DNA (see Fig. 3). We disregard crystal-field distortion of the atomic wavefunctions, so the relevant coupling results from the overlap between the $l = 2$ wavefunctions of this atom and the DNA site orbitals. In the calculation below we assume that this coupling, between the atomic wavefunction $\psi_{l=2,m_l}(\vec{r} - \vec{r}_A)$ centered at $\vec{r}_A$ and a DNA site wavefunction centered at $\vec{r}_n$ is proportional to $\psi_{l=2,m_l}(\vec{r}_n - \vec{r}_A)$. Otherwise, the substrate density of states in the energy range relevant for the tunneling process is assumed constant. Atomic wavefunctions used are hydrogenic wavefunctions for the $n = 5$ (outer gold) shell, calculated with effective atomic number Z=2 to account for screening by inner shell electrons.

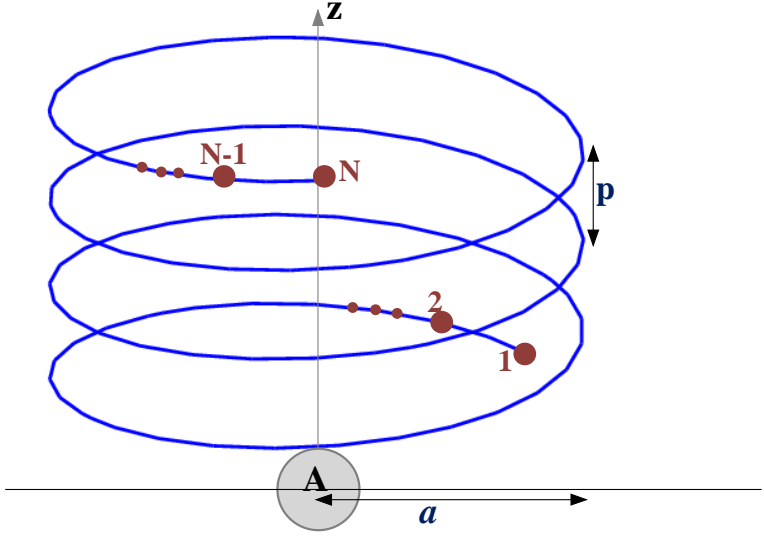

Fig. 3. The DNA-substrate model, with the DNA described as a tight-binding chain while the DNA-substrate coupling is assumed to be dominated by a substrate atom A nearest to the molecule.

To evaluate the transmission probability we need to specify the substrate density of states $\rho$, the position $\vec{r}_1 - \vec{r}_A$ of the helix site 1 nearest to the surface atom (Fig. 3), the self-energy of the helix associated with its coupling to the reservoirs near positions 1 and $N$, and the geometrical and electronic structures of the helix as expressed by the relative position of the helix sites and the intersite coupling. Only the last two intrinsic helix properties affect the resulting spin polarization of the electronic wave injected into the helix, however the overall spin polarization at a detector placed outside the far end of the helix, as expressed by the analog of Eq. (47), also depends on the transmission properties at the two interfaces (see below). In cylindrical coordinates the position of a helix site is written $(z, a, \phi)$, where $a$ is the helix radius. The surface atom is placed at $(z_A = 0, r_A, \phi_A = 0)$ so that $r_A$ measures its distance from the symmetric position on the axis, the first helix site is placed at $(z_1, a, \phi_1 = 0)$ and subsequent sites are positioned so that two nearest neighbors are positioned at $(z, a, \phi)$ and $(z \pm p/N_p, a, \phi \pm 2\pi/N_p)$ so that the nearest neighbor distance is $d_{nn} = \sqrt{2a^2\left(1-\cos\left(2\pi/N_p\right)\right)+\left(p/N_p\right)^2}$. In our calculation we use typical DNA values for the helix: radius, $a = 1\,\text{nm}$, pitch, $p = 3.4\,\text{nm}$ and number of sites per pitch, $N_p = 11$. The intersite coupling is set to $V = 1$ and is used in what follows as our unit of energy.

Because of the Kramers degeneracy, substrate states that belong to a given $j$ band must appear as degenerate pairs $\{\pm m_j\}$. From $m_j = m_l + m_s$, it follows that for each substrate state with a given $m_s$ parent there is a substrate state of the same energy with the opposite, $-m_s$, parent. Therefore, if the transmission process is by itself spin independent, without S-L coupling both spin orientations will be expressed in the transmitted flux in equal amounts. As seen in Section 3, spin selectivity can be affected in the transmission process by the combined effect of (a) dependence of the transmission on the orbital motion and (b) the spin-orbit coupling in the substrate or at the substrate surface.

The calculation proceeds by rewriting Eqs. (44b) and (45) so as to take into account the actual selection rules. For our problem the operators of interest are the orbital angular momentum $\hat{L}$, the spin $\hat{S}$ and the total angular momentum $\hat{J}$ as well as their projections $\hat{L}_z, \hat{S}_z$ and $\hat{J}_z$. The corresponding quantum numbers are $s = 1/2$ and $j = 5/2$ and $l = 2$ as determined by our assumption concerning the incoming electrons. Also, the spin projection, $m_s$, is determined by the final measurement that checks whether $m_s$ is +1/2 or -1/2. The expressions for the transmission function equivalent to Eqs. (45) and (44b) are

$$T_{2\,D_{5/2} \to m_s}(E) = \sum_n \sum_{n'} \langle n|\Gamma^{(m_s)}(E)|n'\rangle \langle n'|G^\dagger(E)|N\rangle \Gamma_N(E)\langle N|G(E)|n\rangle \qquad (65)$$

$$\left(\Gamma^{(m_s)}\right)_{n,n'} \equiv 2\pi\rho \sum_{m_j}\sum_{m_l}\sum_{m_l'} \left\langle 2,m_l,\frac{1}{2},m_s\middle|\frac{5}{2},m_j\right\rangle \left\langle 2,m_l',\frac{1}{2},m_s\middle|\frac{5}{2},m_j\right\rangle^* V_{n,m_l} V_{m_l',n'} \qquad (66)$$

where $\langle l, m_l, s, m_s | j, m_j \rangle$ are Clebsch-Gordan coefficients. Since these coefficients vanish unless $m_j = m_l + m_s$, Eq. (66) can be simplified. We get

$$\left(\Gamma^{(m_s)}\right)_{n,n'} \equiv 2\pi\rho \sum_{m_l} \left|\left\langle 2,m_l,\frac{1}{2},m_s\middle|\frac{5}{2},m_l+m_s\right\rangle\right|^2 V_{n,m_l} V_{m_l,n'} \qquad (67)$$

In obtaining this result we have assumed that all states of the $^2d_{5/2}$ sub-band of gold contribute equally to the transmission. Other models could be considered. For example, a more careful study of the density of states of the $^2d_{5/2}$ sub-band for gold shows peaks in the density of states that arise from Stark splitting of the different $|m_j|$ atomic states by the crystal electric field.[27] The $\pm m_j$ states with the highest energy fill the energy interval within a depth of 2.8 eV below the Fermi level. If we assume that only this group of $m_j$ states contributes to the photoemission signal, the calculation described above will be modified. For example, if these states correspond to $m_j = \pm \frac{5}{2}$, that is, only these values of $m_j$ contribute to Eq. (67), this equation becomes

$$\left(\Gamma^{(\pm 1/2)}\right)_{n,n'} \equiv 2\pi\rho \left|\left\langle 2,\pm 2, \frac{1}{2}, \pm 1/2 \middle| \frac{5}{2}, 5/2 \right\rangle\right|^2 V_{n,m_l=\pm 2} V_{m_l=\pm 2,n'} \tag{68}$$

The model should be supplemented by the self-energies that account for the coupling of the helix to its environment. For the self-energy at the far end (site $N$) of the molecular helix we consider two models. In one, we take a completely transparent boundary, in effect assuming that the helix extends to infinite length, by associating with end site the exact self-energy of a tight binding lattice,

$$\Sigma_N(\varepsilon) = \frac{\varepsilon - \varepsilon_0 - \sqrt{(\varepsilon - \varepsilon_0)^2 - 4V^2}}{2} \equiv \Lambda_N(\varepsilon) - \frac{i}{2}\Gamma_N(\varepsilon) \tag{69}$$

where $\varepsilon_0$ and $V$ are the site energy and nearest neighbor coupling of the molecular tight-binding model. In the other model, we assume that the space outside site $N$ is characterized by a wide-band spectrum, and associate with this site a constant damping rate $\Gamma_N$ (i.e. $\Sigma_N = -(i/2)\Gamma_N$). On the surface side, one contribution to the self-energy comes from the coupling to the surface atom that dominates the electron injection. For a given $m_s (= \pm 1/2)$ state this is

$$\left(\Sigma_A^{(m_s)}\right)_{n,n'} = -(i/2)\left(\Gamma^{(m_s)}\right)_{n,n'} \tag{70}$$

In addition, we assign a self-energy to site 1 of the helix that accounts for electron flux losses all other available states of the substrate. For this self-energy, $\Sigma_1$ we take again one of the two models used for $\Sigma_N$, that is, either the tight binding expression (69) or a constant $-(i/2)\Gamma_1$. Finally, the Green functions that appear in Eq. (65) are obtained by inverting the Hamiltonian matrix of the helix, including the relevant self-energies

$$\mathbf{G}^r(\varepsilon) = \left[\varepsilon\mathbf{I} - \mathbf{H} - \Sigma_1(\varepsilon) - \Sigma_N(\varepsilon) - \Sigma_A\right]^{-1} \tag{71}$$

where $\mathbf{H}$ is the nearest-neighbor tight-binding Hamiltonian of the helix.

Results of these calculations are shown in Figures 4a-c, which show the asymmetry factor

$$\mathcal{A}(E) = \frac{\mathcal{T}_{2_{D_{5/2} \to m_s = 1/2}}(E) - \mathcal{T}_{2_{D_{5/2} \to m_s = -1/2}}(E)}{\mathcal{T}_{2_{D_{5/2} \to m_s = 1/2}}(E) + \mathcal{T}_{2_{D_{5/2} \to m_s = -1/2}}(E)} \tag{72}$$

as a function of the transmission energy. Here $m_s = +1/2$ corresponds to spin projection pointing towards the positive z direction, that is, away from the surface. Figure 4a shows the asymmetry factor in a model where $\Sigma_1$ and $\Sigma_N$ are both given by Eq. (69), while Fig. 4b show similar results for the model with $\Sigma_j = -(i/2)\Gamma_j$ with $\Gamma_j = 2$ for $j = 1, N$. The full (blue) line is the result for a calculation based on Eq. (67), that is, assuming that all $m_j$ states of the $j = 5/2$ band contribute equally, while the dashed (green) line corresponds to Eq. (68) that singles out the contribution of the $m_j = \pm 5/2$ states. In these calculations the substrate atom A is placed on the helix axis, in cartesian position $(x_A, y_A, z_A) = (0., 0., -0.1)$ nm, while the position of the nearest helix site is $(x_1, y_1, z_1) = (0., 1., 0.)$ nm. Fig. 4c shows the effect of breaking this axial symmetry, taking $(x_A, y_A, z_A) = (0., 0.5, -0.1)$ nm. The following observations should be pointed out:

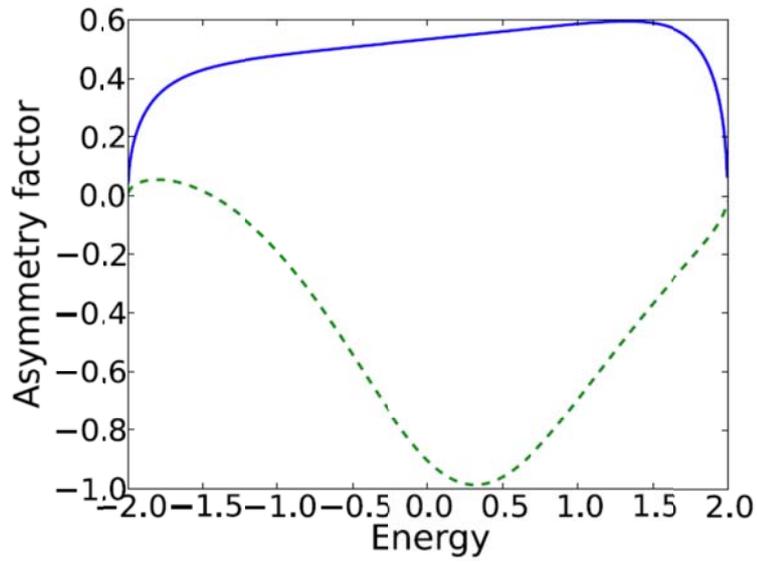

Fig. 4a. The asymmetry factor, Eq. (72), plotted against the transmission energy (in units of the intersite coupling on the helix. Note that the transmission vanishes at the band edges, $E = \pm 2$). The self-energies at sites 1 and $N$ are taken from Eq. (69). The full (blue) line shows the result of a calculation that takes all contributions associated with the $j = 5/2$ substrate band, while the dashed (green) line corresponds to the case where only $m_j = \pm 5/2$ contributes. These results do not depend on the helix length.

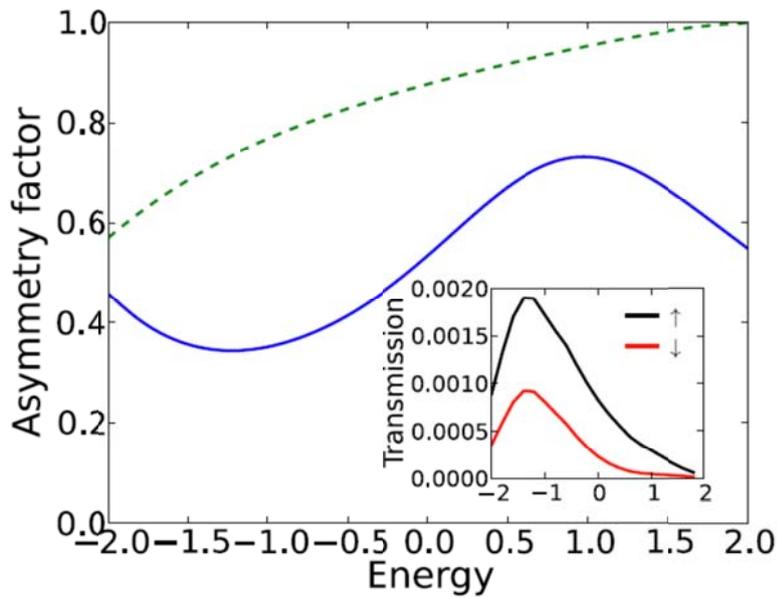

Fig. 4b. Same as 4a, except that the self-energies at sites 1 and $N$ are taken to be $\Sigma_{1,N} = -(i/2)\Gamma$, with $\Gamma = 2$. The helix-length dependence of these results is negligible in the length range (a few pitches) considered. The inset shows the transmission function for the two outgoing spin configurations.

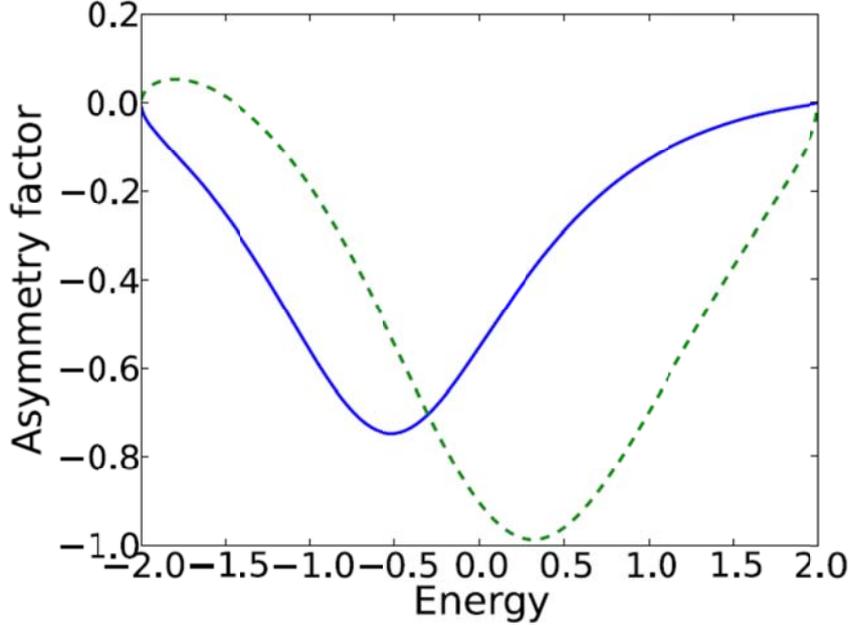

Fig. 4c. Same as Figs 4a,b , exept that the injecting atom is positioned off axis, at $(x_A, y_A, z_A) = (0., 0.5, -0.1)$ nm.

(a) Considerable spin polarizations can be obtained for electron transmission out of a metal substrate through a helical molecule by a mechanism that relies on strong spin-orbit coupling in the substrate together with orbital angular momentum selectivity imposed by the helix. This polarization can be substantial and remains so even when the axial symmetry is broken (Fig. 4c). (b) The magnitude of the effect as well as its sign (positive asymmetry factor implies that the electron spin is preferably polarized in the direction out of the surface) is sensitive to the boundary conditions and the interfacial geometry. We recall that negative asymmetry factor is observed in the photemission experiments,[1] however the present calculation is more appropriate for tunneling transmission,[2] (see Section 5 for treatment of photoemission), for which the sign of the spin polarization cannot be determined.

(c) We have found (not shown) that when $\Gamma_1$ is set to zero, the asymmetry factor, Eq. (72), becomes practically zero. It should be pointed out that the effect of reflection is expected to be less pronounced in pulse experiments if the signal is over before appreciable reflection sets in, see, e.g. Ref. [34].

(d) In the reflectionless case, the length of the helix does not affect the resulting spin polarization. In the presence of reflection (Fig. 4b) the length dependence is still very small for lengths in the range of a few helix pitches. We note that the effect of molecular length observed in the tunneling transmission experiment [2] is not very pronounced above the experimental noise.

We conclude that this simple model of tunneling transmission can account for the observed spin polarization for tunneling out of gold. The computed polarization is positive and essentially independent of molecular length. It is however sensitive to reflections, and it should be kept in mind that reflections by structural irregularities, which are disregarded here, can translate into length dependence. We defer such considerations to future work.

## 5. Induced spin filtering in a scattering model for photoemission through a monolayer of helical molecules

In this section we examine a different mechanism for induced spin filtering by the molecular helix, perhaps better suited to account for over-barrier transmission such as takes place in photoemission. The electron is assumed to have been excited by the light to a free particle state moving in the z (outward, normal to the surface) direction with enough energy to exit. We further assume that elastic collisions with the molecular adsorbate are the primary source for filtering electrons away from the outgoing flux. The calculation is simplified by an additional, rather strong, assumption, that a single collision with a molecular helix makes this electron lost to the detector (the actual process may involve consecutive collisions). Our goal is to determine the cross section for such collision and its dependence on the azimuthal quantum number $m_l$. To this end we start with the Schrodinger equation

$$\left(\nabla^2 + k^2\right)\psi(\vec{r}) = \frac{2m}{\hbar^2} V(\vec{r}) \psi(\vec{r}) \tag{73}$$

where $k^2 = 2mE/\hbar^2$. The relevant solution to Eq. (73) is expressed as a sum of an incident plane wave and a scattered wave

$$\psi(\vec{r}) = e^{ikz} + \psi_s(\vec{r}) \tag{74}$$

The incident wave, a solution of the homogeneous Helmholtz equation, represents the electron emitted by a photoexcited atom. Using a plane wave moving in the *z* direction (normal to the surface) is a choice based on our expectation that such waves are most likely to emerge through the adsorbed molecular layer both because they travel parallel to the molecular chains and because they carry the highest available energy in the exit direction.

In terms of the Green function that satisfies the inhomogeneous Helmholtz equation with a point source,

$$\left(\nabla^2 + k^2\right) G(\vec{r}, \vec{r}\,') = \delta(\vec{r} - \vec{r}\,') \tag{75}$$

the scattered wavefunction satisfies

$$\psi_s(\vec{r}) = \frac{2m}{\hbar^2} \int G(\vec{r}, \vec{r}\,') V(\vec{r}\,') \psi(\vec{r}\,') d\vec{r}\,' \tag{76}$$

and in the first Born approximation

$$\psi_s^{(1B)}(\vec{r}) = \frac{2m}{\hbar^2} \int G(\vec{r}, \vec{r}\,') V(\vec{r}\,') e^{ikz'} d\vec{r}\,' \tag{77}$$

We are interested in the asymptotic form, $r \to \infty$ of this function. To this end we use the asymptotic form of the Green function in cylindrical coordinates $(R, \phi, z)$ (see Appendix D)

$$G(\vec{r}, \vec{r}\,') \xrightarrow{r \to \infty} -\frac{1}{4\pi} \frac{e^{ikr}}{r} e^{-ikz'\cos\theta} \sum_{n=0}^{\infty} \epsilon_n \cos[n(\phi - \phi')](-i)^n J_n(kR'\sin\theta) \tag{78}$$

where $\varepsilon_n$ is given by Eq. (148). Using this in Eq. (77) yields

$$\psi_s^{(1B)}(\vec{r}) = f(\theta, \phi) \frac{e^{ikr}}{r} \tag{79}$$

where the scattering amplitude is given by

$$f(\theta, \phi) = -\frac{m}{2\pi\hbar^2} \sum_{n=0}^{\infty} (-i)^n \epsilon_n \int_{-\infty}^{\infty} dz' \int_0^{\infty} dR' R' \int_0^{2\pi} d\phi' \cos\left(n(\phi - \phi')\right) e^{ikz'(1-\cos\theta)} J_n(kR'\sin\theta) V(\vec{r}\,')$$

$$\tag{80}$$

As a model for the DNA molecule we represent the scattering potential $V(\vec{r}')$ as a helical delta-function potential

$$V(\vec{r}') = V_0 \delta\left(x' - a\cos\left(\frac{2\pi z'}{p}\right)\right) \delta\left(y' - a\sin\left(\frac{2\pi z'}{p}\right)\right) \Theta(z')\Theta(L-z') \quad (81)$$

where $a$ is the radius of the helix, $L$ is its length, $p$ is the pitch and $V_0$ is a constant of dimension energy x length$^2$. In cylindrical coordinates this translates into

$$V(\vec{r}') = \frac{V_0}{R'} \delta(R' - a) \delta\left(\phi' - \frac{2\pi z'}{p}\right) \Theta(z')\Theta(L-z') \quad (82)$$

which, when used with Eq. (80), leads to

$$f(\theta,\phi) = -\frac{mV_0}{2\pi\hbar^2} \sum_{n=0}^{\infty} (-i)^n \epsilon_n \int_0^L dz' \cos\left(n\left(\phi - \frac{2\pi z'}{p}\right)\right) e^{ikz'(1-\cos\theta)} J_n(ka\sin\theta) \quad (83)$$

Evaluating the integral and using the identities $(-i)^{-n} = i^n$ and $J_{-n}(x) = (-)^n J_n(x)$ yields

$$f(\theta,\phi) = \sum_{n=-\infty}^{\infty} f_n(\theta) e^{in\phi} \quad (84)$$

where

$$f_n(\theta) = -\frac{mV_0}{2\pi\hbar^2} (-i)^n J_n(ka\sin\theta) \frac{e^{i\left(k(1-\cos\theta) - \frac{2\pi n}{p}\right)L} - 1}{i\left(k(1-\cos\theta) - \frac{2\pi n}{p}\right)} \quad (85)$$

The differential scattering cross section is

$$\frac{d\sigma}{d\Omega} = |f(\theta,\phi)|^2 = \sum_{n=-\infty}^{\infty} \sum_{n'=-\infty}^{\infty} f_{n'}^*(\theta) f_n(\theta) e^{i(n-n')\phi} \quad (86)$$

and the total scattering cross-section is $\int_0^\pi \sin\theta d\theta \int_0^{2\pi} d\phi |f(\theta,\phi)|^2$. Using $\int_0^{2\pi} d\phi e^{i(n-n')\phi} = 2\pi \delta_{n,n'}$, the total cross-section is obtained in the form

$$\sigma = \sum_{n=-\infty}^{\infty} \sigma_n \quad (87)$$

where

$$\sigma_n = \left(\frac{mV_0}{\pi\hbar^2}\right)^2 2\pi \int_0^\pi d\theta \sin\theta J_n^2(ka\sin\theta) \left(\frac{\sin\left(\left(k(1-\cos\theta)-\frac{2\pi n}{p}\right)\frac{L}{2}\right)}{k(1-\cos\theta)-\frac{2\pi n}{p}}\right)^2 \quad (88)$$

For the set of (positive) $n$ values that satisfy

$$0 < \frac{\pi n}{kp} < 1 \quad (89)$$

the denominator in Eq. (88) can vanish and the partial cross-sections given by Eq. (88) are particularly large. They can be evaluated for large $L$ by using $\int_{-\infty}^{\infty} w^{-2} \sin^2 wL \, dw = \pi L$ to make the approximation

$$\frac{\sin^2(wL)}{w^2} \to L\pi\delta(w) \quad (90)$$

so that, for $n$ in this range

$$\sigma_n = \frac{m^2 V_0^2 L}{\hbar^4 k} J_n^2\left(ka\sqrt{1-\left(1-\frac{2\pi n}{kp}\right)^2}\right) \quad (91)$$

The resonant condition, vanishing of the denominator in Eq. (88) may be given a simple physical interpretation. Consider an electron that is incident along the helix and follows two paths, labeled 1 and 2 in Fig. 5. Path 1 is longer than path 2 by an amount $\Delta l = p - p\cos\theta$. The condition for constructive interference is $\Delta l = n\lambda = 2\pi n/k$. So the resonance condition becomes $k(1-\cos\theta) - 2\pi n/p = 0$, which is precisely the form of the denominator. For those angles which satisfy this condition constructive interference results in strong scattering.

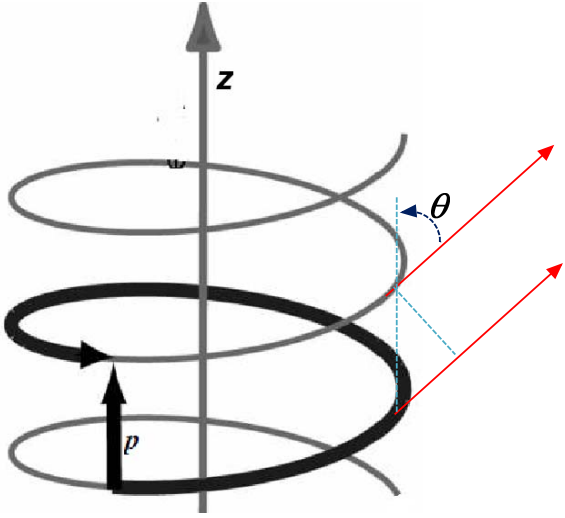

Fig. 5. Diffractive scattering from an helix: In terpretation of Eq. (88).

For *n* outside the range in Eq. (89), including all $n < 0$, the cross section for scattering remains small, and becomes independent of *L* for large *L*. To see this we note that the rapidly oscillating $\sin^2$ function in Eq. (88) can in this case be approximated by it average ½, so Eq. (88) becomes, for *n* outside the range of Eq. (89)

$$\sigma_n \approx \left(\frac{mV_0}{\pi\hbar^2}\right)^2 \pi \int_0^\pi d\theta \sin\theta J_n^2(ka\sin\theta) \left(\frac{1}{k(1-\cos\theta) - \frac{2\pi n}{p}}\right)^2 \quad (92)$$

These observations are confirmed by numerical evaluation of the full expression (88). As an example, the reduced partial cross-section, $\bar{\sigma}_n = \left[\left(\frac{mV_0}{\pi\hbar^2}\right)^2 2\pi\right]^{-1} \sigma_n$, is shown as a function of *L* in Fig. 6, using typical DNA parameters: $a$ = 1.0 nm, $p$ = 3.4 nm and energy $E = \hbar^2 k^2/2m_e = 0.5\,\text{eV}$ ($m_e$ = electron mass). The different modes of *L* dependence in the $n = \pm 1$ cases are clearly shown.

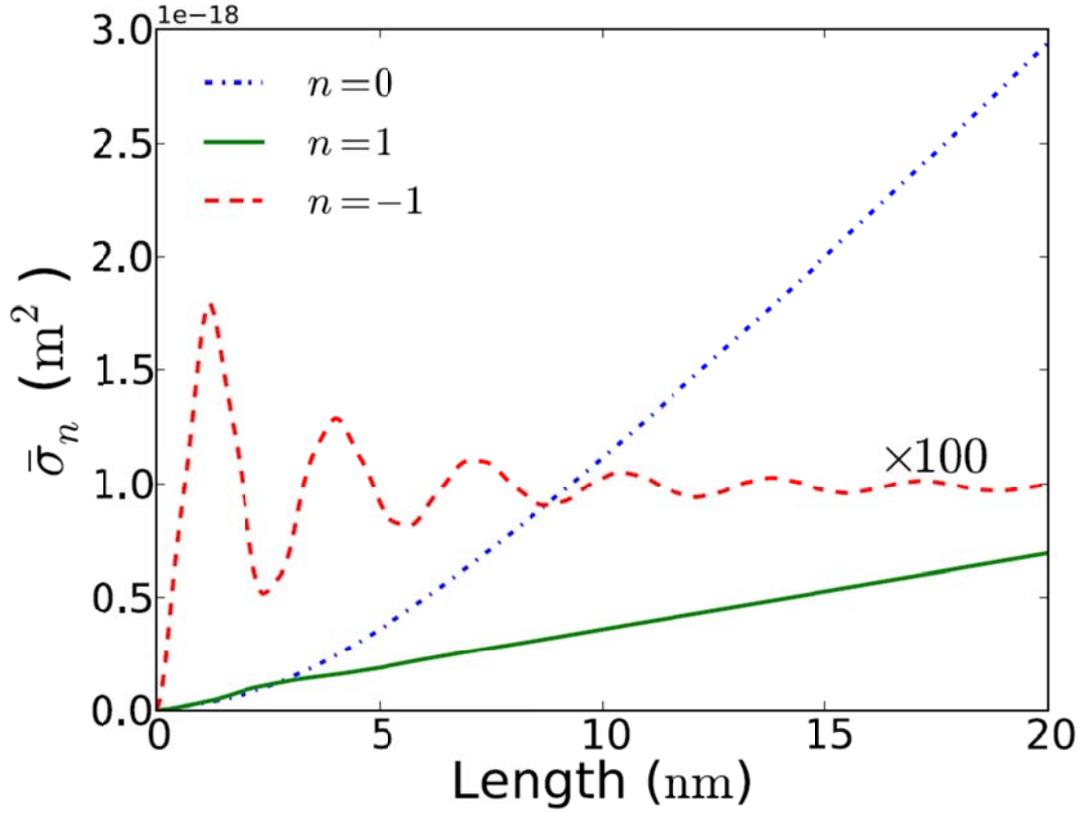

Fig. 6. The reduced cross-section $\bar{\sigma}_n$ plotted against the molecular length $L$ for energy $E$ = 0.5 eV

Thus, in this model only those waves with azimuthal quantum numbers $n$ that satisfy (89) are effectively scattered and therefore filtered out of the transmitted beams, and this effect grows linearly with $L$. Indeed

$$\frac{\sigma_n}{\sigma_{-n}} \approx L \frac{\pi k J_n^2\left(ka\sqrt{1-\left(1-\frac{2\pi n}{kp}\right)^2}\right)}{\left(\int_0^\pi d\theta \sin\theta \, \frac{J_n(ka\sin\theta)}{(1-\cos\theta)+\frac{2\pi n}{kp}}\right)^2} \, ; \quad 0 < \frac{\pi n}{kp} < 1 \qquad (93)$$

The amount of filtering is seen to grow linearly with the molecular length $L$. Finally, as before, the correlation between orbital and spin angular momenta implies that spin selectivity also takes place.

In what follows we make some drastic simplifications in order to estimate the resulting spin filtering effect. First, noting that the azimuthal quantum number $n$ that can contribute to the incoming wave considered above in a given energy region corresponds to the values of the quantum numbers $\pm m_l$ that can be obtained by the photoexcitation of the substrate metal, we assume at the outset that the magnitudes $|m_l|$ of these values fall in the range (89). We further assume that an electron that is scattered by the DNA is lost to the detector. Denote by $N$ the number of DNA molecules absorbed per unit area. The probability that an electron with azimuthal quantum number $m_l$ will pass through molecular layer without scattering is

$$T_{m_l} = 1 - N\sigma_{m_l} = \begin{cases} 1 - N\sigma_{m_l}; & N\sigma_{m_l} < 1 \\ 0 & ; N\sigma_{m_l} \geq 1 \end{cases} \quad (94)$$

In this model it may be possible for the DNA monolayer to be opaque to one value of $m_l$ and yet allow electrons with other values of $m_l$ through.

Consider now the expected degree of spin polarization of electrons photoemitted from gold covered by a monolayer of DNA molecules. For energies close to the photoemission threshold the electrons originate primarily from the relatively narrow d-band of gold and are promoted to the broad p-band conduction band. Those electrons with energies above the vacuum level can pass into the vacuum. Noting again that the d-bands in gold are split into a higher energy $^2D_{5/2}$ and a lower energy $^2D_{3/2}$ band (although there is some overlap between the bands),[27] it will be assumed that the energy of the incident light is sufficiently low that only the $^2D_{5/2}$ band contributes to the photoemitted flux. Since these d-bands are narrow, we will treat them as being atomic-like. Thus, in this simplified picture, photoemission originates from essentially atomic states characterized by total angular momentum quantum number $j = 5/2$ with azimuthal projection quantum numbers $\{m_j\}$ associated with the parent $l = 2$ orbital angular momentum state. (The quantization axis is taken to be normal to the metal surface). The set of $m_j$ states that contribute to the observed photoemission may be further restricted by energy considerations brought about, for example, by crystal fields in the solid.

We examine photoemission by unpolarized incident light and consider separately the contribution from its right- and left-handed components whose handedness is denoted $\mu = \pm 1$.

The selection rules for optically allowed dipole transitions involve the orbital angular momentum $l$ and its azimuthal projection $m_l$. They are

$$\Delta l = \pm 1, \quad \Delta m_l = \mu, \quad \Delta s = \Delta m_s = 0 \tag{95}$$

where $s$ and $m_s$ are the quantum numbers for the spin and its azimuthal projection. To implement these selection rules for an initial $(l=2, j=5/2, m_j)$ state we expand it in terms of the eigenstates $(l, m_l, s, m_s)$ that characterize an atom without spin-orbit coupling, keeping $l=2$. Since for the optical d-band → p-band transition $\Delta l = -1$, it follows that $-(l-1) \leq m_l + \mu \leq l-1$. For $\mu = +1$ we therefore have $-2 \leq m_l \leq 0$, while for $\mu = -1$ we have $0 \leq m_l \leq 2$. The conditional probability to observe a final spin projection $m_s$ for a given $\mu$ is denoted $P_\mu(m_s)$. Since the probability of having either component is $P_\mu = 1/2$, the overall probability to observe a final spin projection using unpolarized light is

$$P(m_s) = P_1(m_s) P_1 + P_{-1}(m_s) P_{-1} = (1/2)\left[ P_1(m_s) + P_{-1}(m_s) \right] \tag{96}$$

We first assume (other assumptions will be considered later) that all the $m_j$ states associated with $(l=2, j=5/2)$ are degenerate with each other, so their relative contribution to the photoemission process is not restricted by their energy. The probability to observe a final spin projection for $\mu = 1$ is then

$$P_1(m_s) = \sum_{m_l = -2}^{0} T_{m_l+1} \left| \sum_{m_j} \left\langle 2, m_l, \frac{1}{2}, m_s \middle| \frac{5}{2}, m_j \right\rangle \right|^2 \tag{97}$$

where $\langle l, m_l, s, m_s | j, m_j \rangle$ are Clebsch-Gordan (CG) coefficients. The sum over $m_j$ can be dropped because $m_j = m_l + m_s$ has to be satisfied (that is, CG=0 unless this is so). Eq. (97) becomes

$$P_1(m_s) = \sum_{m_l = -2}^{0} T_{m_l+1} \left| \left\langle 2, m_l, \frac{1}{2}, m_s \middle| \frac{5}{2}, m_l + m_s \right\rangle \right|^2$$
$$= T_{-1} \left| \left\langle 2, -2, \frac{1}{2}, m_s \middle| \frac{5}{2}, -2 + m_s \right\rangle \right|^2 + T_0 \left| \left\langle 2, -1, \frac{1}{2}, m_s \middle| \frac{5}{2}, -1 + m_s \right\rangle \right|^2 + T_1 \left| \left\langle 2, 0, \frac{1}{2}, m_s \middle| \frac{5}{2}, 0 + m_s \right\rangle \right|^2 \tag{98}$$

Similarly for $\mu = -1$

$$P_{-1}(m_s) = \sum_{m_l=0}^{2} T_{m_l-1} \left| \left\langle 2, m_l, \frac{1}{2}, m_s \middle| \frac{5}{2}, m_l + m_s \right\rangle \right|^2$$

$$= T_{-1} \left| \left\langle 2, 0, \frac{1}{2}, m_s \middle| \frac{5}{2}, 0 + m_s \right\rangle \right|^2 + T_0 \left| \left\langle 2, 1, \frac{1}{2}, m_s \middle| \frac{5}{2}, 1 + m_s \right\rangle \right|^2 + T_1 \left| \left\langle 2, 2, \frac{1}{2}, m_s \middle| \frac{5}{2}, 2 + m_s \right\rangle \right|^2 \tag{99}$$

The needed Clebsch-Gordan coefficients are listed in Appendix E. Using these, Eqs. (98) and (99) lead to

$$P_1\left(\frac{1}{2}\right) = \frac{1}{5}T_{-1} + \frac{2}{5}T_0 + \frac{3}{5}T_1 \tag{100}$$

$$P_1\left(-\frac{1}{2}\right) = T_{-1} + \frac{4}{5}T_0 + \frac{3}{5}T_1 \tag{101}$$

$$P_{-1}\left(\frac{1}{2}\right) = \frac{3}{5}T_{-1} + \frac{4}{5}T_0 + T_1 \tag{102}$$

$$P_{-1}\left(-\frac{1}{2}\right) = \frac{3}{5}T_{-1} + \frac{2}{5}T_0 + \frac{1}{5}T_1 \tag{103}$$

Together the contributions from the two polarization states gives, for spin up

$$P(1/2) = \frac{P_1(1/2) + P_{-1}(1/2)}{2} = \frac{2}{5}T_{-1} + \frac{3}{5}T_0 + \frac{4}{5}T_1 \tag{104}$$

Similarly, for spin down

$$P(-1/2) = \frac{P_1(-1/2) + P_{-1}(-1/2)}{2} = \frac{4}{5}T_{-1} + \frac{3}{5}T_0 + \frac{2}{5}T_1 \tag{105}$$

Finally, the spin polarization asymmetry ratio is

$$\mathcal{A} = \frac{P(1/2) - P(-1/2)}{P(1/2) + P(-1/2)} = \frac{T_1 - T_{-1}}{3(T_{-1} + T_0 + T_1)} \tag{106}$$

From Figure 5 we see that for large $L$, $\sigma_0 \gg \sigma_1 \gg \sigma_{-1}$. If, for the sake of quick estimate, we invoke Eq. (94) to assume that $T_{-1} = 1$ while $T_0 = T_1 = 0$, we get the polarization ratio $\wp = -1/3$, to be compared with the observed polarization $\sim -0.6$.

This rough estimate should be regarded more as an example of what can be estimated from such arguments rather than a theoretical prediction. Other quick estimates may be attempted. For example, if we assume as in Section 4 that only $m_j$ states with the highest energy contribute to the photoemission signal, and if these states correspond to $m_j = \pm\frac{5}{2}$, Eqs. (98) and (99) become

$$P_1(m_s) = T_{-1}\left|\left\langle 2,-2,\frac{1}{2},m_s \bigg| \frac{5}{2},-2+m_s \right\rangle\right|^2 \delta_{m_s,-1/2} \tag{107}$$

$$P_{-1}(m_s) = T_1\left|\left\langle 2,2,\frac{1}{2},m_s \bigg| \frac{5}{2},2+m_s \right\rangle\right|^2 \delta_{m_s,1/2} \tag{108}$$

and Eqs. (100)-(105) are replaced by

$$P_1\left(-\frac{1}{2}\right) = T_{-1}; \quad P_1\left(\frac{1}{2}\right) = 0 \tag{109}$$

$$P_{-1}\left(-\frac{1}{2}\right) = 0 \; ; \quad P_{-1}\left(\frac{1}{2}\right) = T_1 \tag{110}$$

$$P(1/2) = \frac{P_1(1/2)+P_{-1}(1/2)}{2} = \frac{1}{2}T_1 \tag{111}$$

$$P(-1/2) = \frac{P_1(-1/2)+P_{-1}(-1/2)}{2} = \frac{1}{2}T_{-1} \tag{112}$$

The spin polarization for this case is

$$\wp = \frac{P(1/2)-P(-1/2)}{P(1/2)+P(-1/2)} = \frac{T_1 - T_{-1}}{T_1 + T_{-1}} \tag{113}$$

and for $T_1 = 0$; $T_{-1} = 1$ we get $\wp = -1$, that is full polarization towards to surface. Obviously, the observed result, $\wp = -2/3$, can be obtained in intermediate situations.

## 6. Summary and conclusions

Three issues were discussed in this paper: We have first argued that spin orbit coupling induced by electron motion through a helical structure cannot, by itself, account for recent observations of large spin selectivity in photoemission through such structures. Second, and most important, we have introduced the concept of induced selectivity or induced filtering - selectivity in the dynamical evolution of one observable can induce selectivity in another observable that is coupled to it. We have demonstrated such induced filtering in transmission between two reservoirs: one in which two such dynamical variables are coupled and another reservoir where they are not, through a bridge whose transmission properties depend only on the state of one of these variables. Another example is electron photoemission from surfaces characterized by strong spin-orbit coupling using circularly polarized light. Third, we have applied this theoretical framework to the interpretation of recent experimental observations of large spin selectivity in electron photoemission and tunneling process through DNA and other chiral molecules, where at least one of the metals involved is gold or silver - metals characterized by strong spin-orbit coupling. We have studied two models: in one, appropriate for tunneling situations, we have estimated the spin polarizability in the transmission current calculated from the Landauer formula. This model predicts positive spin polarizability in the transmitted current and does not show molecular length dependence of the effect in the absence of dephasing processes. In another, more suitable for over-barrier transmission as in photoemission, we have studied the spin selectivity induced by the orbital angular momentum dependence of electron scattering by helical structures. This model yields negative spin polarization that increases linearly with the helix length in the range studied. In either case we considered only elastic process. It will be of interest to consider the possible consequences of energy losses in future studies, but at first glance it seems that such effects are small in the energy range relevant to current experimental results, that is below the electronic excitation spectrum of DNA. Another consideration for future study is the possibility that the adsorbed helical layer affects the nature of the incident light, perhaps inducing some circular polarization character that is expressed in the photoexcitation process.

Both models considered yield spin polarization of the observed order of magnitude using reasonable parameters for the system geometry and its electronic structure. These results should be regarded as estimates only and should be repeated with more detailed structural data for the specific systems used in the experiments. In particular, we have used the bulk electronic structure of gold as a guide for our arguments, while, obviously, the surface electronic structure should also be considered in rigorous calculations. While our results seem to be in accord with published experimental results on gold and silver, recent observation of considerable spin polarization in the photoemission from bacteriorhodopsin covered aluminum bring up new questions. By itself, aluminum is a low spin-orbit coupling material, so the mechanism discussed in this paper can be relevant only provided such coupling is caused at the molecule-metal interface by the interfacial built-in potential. Obviously this may also be an indication that another mechanism, yet unknown, is at play. These issues will be subjects of future studies.

**Appendix A. Perturbative estimate of spin rotation.**

The initial conditions for the dynamical Pauli matrices are the usual Pauli spin matrices

$$\vec{\sigma}(0) = \begin{pmatrix} 0 & 1 \\ 1 & 0 \end{pmatrix}\hat{i} + \begin{pmatrix} 0 & -i \\ i & 0 \end{pmatrix}\hat{j} + \begin{pmatrix} 1 & 0 \\ 0 & -1 \end{pmatrix}\hat{k} \qquad (114)$$

Assuming that the speed $v$ varies in the range $10^5$-$10^7$ m/s, the parameters $g$ and $b$ of Eq. (11) assume values in the ranges $g \sim 3 \times 10^{-4}$ - $3 \times 10^{-2}$ and $b \sim 1. \times 10^{-7}$ - $1. \times 10^{-3}$. Note that $b$ is roughly the same size as $g^2$. Thus both parameters are small in magnitude and this suggests that a perturbation solution of the equations of motion would suffice.

To lowest order in both $g$ and $b$ the solutions of Eqs. (12) are

$$\sigma_x(\theta) \approx \begin{pmatrix} -b\sin\theta & 1+2ig^2\theta \\ 1-2ig^2\theta & b\sin\theta \end{pmatrix}$$

$$\sigma_y(\theta) \approx \begin{pmatrix} b(\cos\theta-1) & -i+2g^2\theta \\ i+2g^2\theta & b(1-\cos\theta) \end{pmatrix} \qquad (115)$$

$$\sigma_z(\theta) \approx \begin{pmatrix} 1 & ib(e^{-i\theta}-1) \\ -ib(e^{i\theta}-1) & -1 \end{pmatrix}$$

If the initial state is one of positive spin projection, that is $\psi = \begin{pmatrix} 1 \\ 0 \end{pmatrix}$, the expectation values of the Pauli spin matrix components are $\langle \sigma_x(0) \rangle = 0$, $\langle \sigma_y(0) \rangle = 0$, $\langle \sigma_z(0) \rangle = 1$.

After traversing some length of the helix the expectation values become

$$\langle \sigma_x(\theta) \rangle \approx -b\sin\theta, \quad \langle \sigma_y(\theta) \rangle \approx b(\cos\theta-1), \quad \langle \sigma_z(\theta) \rangle \approx 1 \qquad (116)$$

If the helix consists of $N$ turns (where $N$ need not be an integer) then $\theta = 2\pi N$. Since $b$ is a small number the spin projection does not change much from its starting value. In this approximation, when $N$ is an integer the expectation values return to their original values.

**Appendix B. Spin flip by spin-orbit scattering off a helical potential**

Here we start from Eqs. (17)-(24) and derive Eq. (25). First note that

$$\vec{\sigma} \cdot \hat{k} \times \nabla V = \hat{k} \cdot \nabla V \times \sigma = \sigma_y \partial_x V - \sigma_x \partial_y V \tag{117}$$

Thus, Eq. (24) becomes

$$H_{so} = -\frac{\hbar v}{4mc^2} \begin{pmatrix} 0 & -\partial_y V - i\partial_x V \\ -\partial_y V + i\partial_x V & 0 \end{pmatrix} \tag{118}$$

The off-diagonal matrix elements connect spin-up states, $\begin{pmatrix} 1 \\ 0 \end{pmatrix}$, and spin-down states, $\begin{pmatrix} 0 \\ 1 \end{pmatrix}$, and causes spin flipping. Carrying out the derivatives gives

$$\begin{aligned}\partial_x V = &V_0 cos\varphi \delta'(xcos\varphi + ysin\varphi - a)\delta(-xsin\varphi + ycos\varphi) \\ &- V_0 sin\varphi \delta(xcos\varphi + ysin\varphi - a)\delta'(-xsin\varphi + ycos\varphi)\end{aligned} \tag{119}$$

and

$$\begin{aligned}\partial_y V = &V_0 sin\varphi \delta'(xcos\varphi + ysin\varphi - a)\delta(-xsin\varphi + ycos\varphi) \\ &+ V_0 cos\varphi \delta(xcos\varphi + ysin\varphi - a)\delta'(-xsin\varphi + ycos\varphi)\end{aligned} \tag{120}$$

As noted in the main text, the effect of the helix is on the *xy*-motion of the electron. We next use time-dependent perturbation theory (first order) to calculate the amplitude for making a transition from an initial state $(\vec{k}_\perp, s)$ to a final state $(\vec{k}'_\perp, s')$, i.e., $\psi_i = \chi_s e^{i\vec{k}_\perp \cdot \vec{R}}/\sqrt{A}$ and $\psi_f = \chi_{s'} e^{i\vec{k}'_\perp \cdot \vec{R}}/\sqrt{A}$, where $\vec{k}_\perp$ corresponds to motion in the *xy*-plane, $A$ is the normalization area and $\chi_s$ is the spin vector. Later it will be assumed that the scattering is elastic, *i.e.*, it changes only the direction of $\vec{k}_\perp$ but not its magnitude. (Affecting the spin is also not an energetic issue in the absence of a magnetic field).

In what follows we disregard scattering by $V(x,y,t)$ and only take into account the magnetic coupling, that is, consider scattering by $H_{so}$ only.[46] The transition amplitude is

$$c_{\vec{k}'_\perp, s'} = -\frac{i}{\hbar} \int_0^T e^{-\frac{i}{\hbar}(E-E')t} \langle \vec{k}'_\perp, s'|H_{so}|\vec{k}_\perp, s \rangle dt \tag{121}$$

where *T* is the transit time. Thus, introducing the wave-vector transfer $\vec{q} = \vec{k}_\perp - \vec{k}'_\perp$, we get

$$\langle \vec{k}'_\perp, s'|H_{so}|\vec{k}_\perp, s \rangle = -\frac{\hbar v}{4mc^2} \frac{1}{A} \int d^2r e^{i\vec{q}\cdot\vec{r}_\perp} \chi_{s'}^+ \begin{pmatrix} 0 & -i(\partial_x - i\partial_y)V \\ i(\partial_x + i\partial_y)V & 0 \end{pmatrix} \chi_s \tag{122}$$

Note that

$$(\partial_x - i\partial_y)V = V_0 e^{-i\varphi}\left[\delta'(x'-a)\delta(y') - i\delta(x'-a)\delta'(y')\right] \tag{123}$$

and

$$(\partial_x + i\partial_y)V = V_0 e^{i\varphi}\left[\delta'(x'-a)\delta(y') + i\delta(x'-a)\delta'(y')\right] \tag{124}$$

It is convenient to introduce rotated wave-vector transfer components

$$q'_x = q_x\cos\varphi + q_y\sin\varphi \tag{125}$$

and

$$q'_y = -q_x\sin\varphi + q_y\cos\varphi \tag{126}$$

and to recall that

$$x' = x\cos\varphi + y\sin\varphi \tag{127}$$

and

$$y' = -x\sin\varphi + y\cos\varphi \tag{128}$$

Note that this is a time-dependent transformation. The variables $x$ and $y$ are defined in a fixed coordinate system, while $x'$ and $y'$ are defined according to a coordinate system that rotates in time. Then

$$\vec{q}'\cdot\vec{r}' = \vec{q}\cdot\vec{r}_\perp \tag{129}$$

and

$$d^2r' = d^2r \tag{130}$$

Integration by parts yields

$$\int d^2r'_\perp\, e^{i\vec{q}'\cdot\vec{r}'_\perp}\delta'(x'-a)\delta(y) = -iq'_x\, e^{iq'_x a} \tag{131}$$

and

$$\int d^2r'_\perp\, e^{i\vec{q}'\cdot\vec{r}'_\perp}\delta(x'-a)\delta'(y) = -iq'_y\, e^{iq'_x a} \tag{132}$$

So

$$\langle \vec{k}'_\perp, s'|H_s|\vec{k}_\perp, s\rangle = -\frac{\hbar v V_0}{4mc^2}\frac{1}{A}e^{iq'_x a}\chi^\dagger_{s'}\begin{pmatrix} 0 & -q'_- e^{-i\varphi} \\ q'_+ e^{i\varphi} & 0 \end{pmatrix}\chi_s$$
$$= -\frac{\hbar v V_0}{4mc^2}\frac{1}{A}e^{i(q_x\cos\varphi + q_y\sin\varphi)a}\chi^\dagger_{s'}\begin{pmatrix} 0 & -q_- \\ q_+ & 0 \end{pmatrix}\chi_s \tag{133}$$

where $q_\pm = q_x \pm iq_y$. The transition amplitude becomes

$$c_{\vec{k}'_\perp,s'} = \frac{ivV_0}{4mc^2}\frac{1}{A}\int_0^T e^{-\frac{i}{\hbar}(E-E')t} e^{i(q_x cos\Omega t + q_y sin\Omega t)a} \chi_{s'}^\dagger \begin{pmatrix} 0 & -q_- \\ q_+ & 0 \end{pmatrix} \chi_s dt \qquad (134)$$

and $\Omega = \frac{2\pi v}{p}$. It makes sense to take $E' = E$ since we are interested mainly in processes that affect spin, not orbital motion. This assumes that changing the spin did not affect the energy (i.e., recall that $k' = k$, whereas there is direction change $\vec{q} = \vec{k}_\perp - \vec{k}'_\perp$).

Let the length of the helix be $L = Np$, where $N$ is the length in units of the pitch (which need not be an integer). The relevant transit time is $T = L/v$, so

$$c_{\vec{k}'_\perp,s'} = \frac{ivV_0}{4mc^2}\frac{1}{A}\int_0^{Np/v} e^{i(q_x cos\Omega t + q_y sin\Omega t)a} \chi_{s'}^\dagger \begin{pmatrix} 0 & -q_- \\ q_+ & 0 \end{pmatrix} \chi_s dt \qquad (135)$$

Introduce polar coordinates $(q, \theta)$ in place of the Cartesian coordinates $(q_x, q_y)$ so

$$c_{\vec{k}'_\perp,s'} = \frac{ivV_0}{4mc^2}\frac{p}{2\pi v}\frac{1}{A}\int_0^{2\pi N} e^{iqacos(\varphi-\theta)} \chi_{s'}^\dagger \begin{pmatrix} 0 & -q_- \\ q_+ & 0 \end{pmatrix} \chi_s d\varphi \qquad (136)$$

We can, without loss of generality, take $\theta = 0$. This simply means that we define the origin of the cylindrical angle $\varphi$ by the direction of $\vec{q}$. This leads to Eq. (25).

Finally we note that the introduction of a second helix does not change the result by very much. The potential may be written as

$$\begin{aligned}V(x,y) &= V_0\delta(xcos\varphi + ysin\varphi - a)\delta(-xsin\varphi + ycos\varphi) \\ &+ V_0\delta(xcos\varphi' + ysin\varphi' - a)\delta(-xsin\varphi' + ycos\varphi')\end{aligned} \qquad (137)$$

Where $\varphi' = \varphi - \delta$ and $\delta$ is an offset angle distinguishing the second helix from the first. The amplitudes turn out to be just twice what they were before for a single helix.

**Appendix C. Derivation of Eqs. (58)-(60)**

The expansion coefficients $A$ and $B$ in (53) and (55) can be found by matching the wave functions at the surface $z = a$. The continuity of the total wave function at this surface can be expressed in terms of of the radial distance from the $z$-azis, $R$ as

$$\sum_{m_l,m_s} \langle lm_l sm_s | jm_j \rangle v_{j,l}(\sqrt{R^2+a^2}) Y_{l,m_l}(\Theta,0) e^{im_l\varphi} \chi^S_{m_s}$$

$$+ \sum_{m_l,m_s} \int_0^\infty B_{m_l,m_s}(Q) J_{m_l}(QR) e^{-iq'a+im_l\varphi} \chi^S_{m_s} dQ \qquad (138)$$

$$= \sum_{m_l,m_s} \int_0^\infty A_{m_l,m_s}(Q) J_{m_l}(QR) e^{iqa+im_l\varphi} \chi^S_{m_s} dQ$$

where $\Theta = cos^{-1}\left(\dfrac{a}{\sqrt{R^2+a^2}}\right)$. Similarly, the component of the gradient of the wave function in the direction normal to the surface must be continuous. Using

$$\frac{\partial}{\partial z} = cos\theta \frac{\partial}{\partial r} - \frac{sin\theta}{r}\frac{\partial}{\partial \theta} \qquad (139)$$

leads to

$$\sum_{m_l,m_s} \langle lm_l sm_s | jm_j \rangle \left[ cos\theta \frac{\partial}{\partial r} - \frac{sin\theta}{r}\frac{\partial}{\partial \theta} \right] v_{j,l}(r) Y_{l,m_l}(\theta,0) e^{im_l\varphi} \chi^S_{m_s}$$

$$+ \sum_{m_l,m_s} \int_0^\infty B_{m_l,m_s}(Q) J_{m_l}(QR)(-iq') e^{-iq'a+im_l\varphi} \chi^S_{m_s} dQ \qquad (140)$$

$$= \sum_{m_l,m_s} \int_0^\infty A_{m_l,m_s}(Q) J_{m_l}(QR)(iq) e^{iqa+im_l\varphi} \chi^S_{m_s} dQ$$

where $r = \sqrt{R^2+a^2}$ and $\theta = \Theta$. It follows that

$$\int_0^\infty A_{m_l,m_s}(Q) J_{m_l}(QR) e^{iqa} dQ$$
$$= \langle lm_l sm_s | jm_j \rangle v_{j,l}\left(\sqrt{R^2+a^2}\right) Y_{l,m_l}(\Theta,0) + \int_0^\infty B_{m_l,m_s}(Q) J_{m_l}(QR) e^{-iq'a} dQ \qquad (141)$$

and

$$\int_0^\infty A_{m_l,m_s}(Q) J_{m_l}(QR) iq e^{iqa} dQ$$
$$= \langle lm_l sm_s | jm_j \rangle \left[ cos\theta \frac{\partial}{\partial r} - \frac{sin\theta}{r}\frac{\partial}{\partial \theta} \right] v_{j,l}(r) Y_{l,m_l}(\theta,0) \qquad (142)$$
$$+ \int_0^\infty B_{m_l,m_s}(Q) J_{m_l}(QR)(-iq') e^{-iq'a} dQ$$

Multiplying Eq. (141) through by the Bessel function $RJ_{m_l}(Q'R)$, integrating over $R$ using the relation

$$\int_0^\infty J_{m_l}(QR) J_{m_l}(Q'R) R dR = \frac{1}{Q}\delta(Q'-Q) \tag{143}$$

leads to

$$A_{m_l,m_s}(Q)e^{iqa}$$
$$= Q\langle l m_l s m_s | j m_j \rangle \int_0^\infty J_{m_l}(QR) v_{j,l}\left(\sqrt{R^2+a^2}\right) Y_{l,m_l}(\Theta,0) R dR + B_{m_l,m_s}(Q)e^{-iq'a} \tag{144}$$

Similarly from (142),

$$iqA_{m_l,m_s}(Q)e^{iqa}$$
$$= Q\langle l m_l s m_s | j m_j \rangle \int_0^\infty J_{m_l}(QR)\left[\cos\theta \frac{\partial}{\partial r} - \frac{\sin\theta}{r}\frac{\partial}{\partial \theta}\right] v_{j,l}(r) Y_{l,m_l}(\theta,0) R dR \tag{145}$$
$$- iq' B_{m_l,m_s}(Q) e^{-iq'a}$$

Solving Eqs. (144) and (145) for the coefficients $A_{m_l,m_s}(Q)$ and $B_{m_l,m_s}(Q)$ yields the results (58)-(60) for the $A$ and $B$ coefficients.

**Appendix D. The asymptotic Green function, Eq. (78)**

In what follows we will use the following expression for the Green function in cylindrical coordinates $(R,\phi,z)$ [47]

$$G(\vec{r},\vec{r}') = -\frac{1}{4\pi}\sum_{n=0}^\infty \epsilon_n G_H^n(k,R,R',z-z')\cos[n(\phi-\phi')] \tag{146}$$

where

$$G_H^n(k,R,R',z-z') = \frac{1}{\pi}\int_0^\pi \frac{e^{ik\sqrt{R^2+R'^2+(z-z')^2-2RR'\cos(\psi)}}}{\sqrt{R^2+R'^2+(z-z')^2-2RR'\cos(\psi)}}\cos(n\psi)d\psi \tag{147}$$

and

$$\epsilon_n = \begin{cases} 1 & \text{if } n=0 \\ 2 & \text{if } n>0 \end{cases} \tag{148}$$

For the scattering process we are interested in the asymptotic form of the Green function for large $R$ and $|z|$. If $V(\vec{r}')$ is localized in space, the values of $R'$ and $|z'|$ remain bounded and we may expand

$$\sqrt{R^2 + R'^2 + (z-z')^2 - 2RR'\cos(\psi)} \approx r\left(1 - \frac{zz' + RR'\cos\psi}{r^2}\right) = r - z'\cos\theta - R'\sin\theta\cos\psi \quad (149)$$

where $r = \sqrt{R^2 + z^2}$, $\cos\theta = \frac{z}{r}$ and $\sin\theta = \frac{R}{r}$. Note that $(r, \theta, \phi)$ are the spherical coordinates of the point $(R, z, \phi)$ expressed in cylindrical coordinates. From Eqs. (147) and (149) we get

$$\begin{aligned}G_H^n(k, R, R', z-z') &\approx \frac{1}{\pi} \frac{e^{ikr}}{r} \int_0^\pi \cos(n\psi) e^{-ikz'\cos\theta - ikR'\sin\theta\cos\psi} d\psi \\ &= (-i)^n \frac{e^{ikr}}{r} e^{-ikz'\cos\theta} J_n(kR'\sin\theta)\end{aligned} \quad (150)$$

where we have used the integral representation of the Bessel function

$$J_n(z) = \frac{i^n}{\pi} \int_0^\pi e^{-iz\cos\psi} \cos(n\psi) d\psi \quad (151)$$

Using this in (146), the asymptotic Green function is obtained in the form (78).

## Appendix E. Relevant Clebsch-Gordan coefficients

The following table summarizes the Clebsch-Gordan coefficients needed for the evaluation of the scattering cross-sections, Eqs. (98) and (99)

| $l$ | $m_l$ | $s$ | $m_s$ | $j$ | $m_j$ | $\langle l, m_l, s, m_s | jm \rangle$ |
|---|---|---|---|---|---|---|
| 2 | -2 | 1/2 | 1/2 | 5/2 | -3/2 | $\sqrt{1/5}$ |
| 2 | -1 | 1/2 | 1/2 | 5/2 | -1/2 | $\sqrt{2/5}$ |
| 2 | 0 | 1/2 | 1/2 | 5/2 | 1/2 | $\sqrt{3/5}$ |
| 2 | 1 | 1/2 | 1/2 | 5/2 | 3/2 | $\sqrt{4/5}$ |
| 2 | 2 | 1/2 | 1/2 | 5/2 | 5/2 | 1 |
| 2 | -2 | 1/2 | -1/2 | 5/2 | -5/2 | 1 |
| 2 | -1 | 1/2 | -1/2 | 5/2 | -3/2 | $\sqrt{4/5}$ |
| 2 | 0 | 1/2 | -1/2 | 5/2 | -1/2 | $\sqrt{3/5}$ |

| 2 | 1 | 1/2 | -1/2 | 5/2 | 1/2 | $\sqrt{2/5}$ |
| 2 | 2 | 1/2 | -1/2 | 5/2 | 3/2 | $\sqrt{1/5}$ |

**Acknowledgements**. We thank Ron Naaman, Rafael Gutierrez and Guido Burkard for helpful discussions, and R. Naaman for disclosing experimental data prior to publication. The research of A. N. is supported by the Israel Science Foundation, the Israel-US Binational Science Foundation and the European Research Council under the European Union's Seventh Framework Program (FP7/2007-2013; ERC grant agreement no. 226628). K. K. acknowledges support from the Villum Kann Rasmussen Foundation.

[42]    Note that in a crystalline solid the degeneracy of the radial wave function will be further lifted by the crystal electric field, so $v_{j,l}(r) \rightarrow v_{j,l,|m_l|}(r)$ ).

[43]    Such approximation is often used in treating photoemission by taking matrix elements of the dipole operator between an initial atomic state and a plane wave (a free wave function rather than a Coulomb wave).

[44]    In particular, we are using the band structure of bulk gold as a guide in developing our argument, keeping in mind that the actual behavior is determined largely by the surface electronic structure.

[45]    By using a tight binding model without specifically addressing the site orbitals, we are essentially disregarding the effect of curvature of the electronic path; the helical structure enters only through the relative positions of the molecular sites relative to the surface atom. The curvature effect is taken into account in the estimate of the contribution of spin-orbit coupling in

Sect. 2 (see also Refs. [22] and [33]), but it appears unimportant for the discussion in the present section.

[46]    We can, if necessary, go to second order and consider the effects of both V and Hso (we consider the effect of these potentials on a free particle motion along the z-axis).

[47]    J. T. Conway, and H. S. Cohl, Z. Angew. Math. Phys. 61 (2010).